\documentclass[a4paper,fleqn,usenatbib]{mnras}
\usepackage{mathrsfs}
\usepackage{color}
\usepackage{mathptmx}

\usepackage[T1]{fontenc}
\usepackage{ae,aecompl}
\usepackage{mathrsfs}

\usepackage{graphicx}	
\usepackage{amsmath}	
\usepackage{amssymb}	
\usepackage{csquotes}   

\DeclareMathAlphabet{\pazocal}{OMS}{zplm}{m}{n}

\def\Lya{Ly$\rm{\alpha}$~} 
\def\Lyb{Ly$\rm{\beta}$~} 
\def\z5{$z\simeq5$\,} 
 
\def\kmsi{$\mathrm{km^{-1}\,s}\,$}

\def\evp{$[\mathrm{eV\,m}_\mathrm{p}^{-1}]\,$}

\def\uo{$u_{0}\,$}
\def\logto{$\mathrm{log}T_0\,$}
\def\taueff{$\tau_{\rm{eff}}\,$}
\def\zre{$z_{\rm{re}}\,$}
\newcommand{\comment}[1]{} 

\def\HI{\hbox{H$\,\rm \scriptstyle I\ $}}
\def\HII{\hbox{H$\,\rm \scriptstyle II\ $}} 

\def\HeII{\hbox{He$\,\rm \scriptstyle II\ $}}
 
\def\CIV{\hbox{C$\,\rm \scriptstyle IV$}}
\def\SiIV{\hbox{Si$\,\rm \scriptstyle IV$}}
\def\MgII{\hbox{Mg$\,\rm \scriptstyle II$}} 

\title[The thermal history during reionisation]{Inferring the IGM thermal history during reionisation with the Lyman-$\alpha$ forest power spectrum at redshift \boldmath{$z \simeq 5$}} \author[F. Nasir, J.S. Bolton \& G.D. Becker]{Fahad Nasir,$^{1}$\thanks{E-mail: ppxfn@nottingham.ac.uk} James S. Bolton,$^{1}$\thanks{E-mail:james.bolton@nottingham.ac.uk} \& George D. Becker $^{2}$\thanks{E-mail: george.becker@ucr.edu}\\ $^{1}$School of Physics and Astronomy, University of Nottingham, University Park, Nottingham, NG7 2RD, UK\\$^{2}$Department of Physics \& Astronomy, University of California, Riverside, 900 University Avenue, Riverside, CA 92521, USA}

\begin{document}
\label{firstpage}
\pagerange{\pageref{firstpage}--\pageref{lastpage}}
\maketitle

\begin{abstract}
We use cosmological hydrodynamical simulations to assess the
feasibility of constraining the thermal history of the intergalactic
medium during reionisation with the \Lya forest at \z5. The
  integrated thermal history has a measureable impact on the
transmitted flux power spectrum that can be isolated from Doppler
broadening at this redshift.  We parameterise this using the
cumulative energy per proton, $u_0$, deposited into a gas parcel at
the mean background density, a quantity that is tightly linked with
the gas density power spectrum in the simulations. We construct mock
observations of the line of sight \Lya forest power spectrum and use a
Markov Chain Monte Carlo approach to recover $u_{0}$ at redshifts $5
\la z \la 12$. A statistical uncertainty of $\sim 20$ per cent is
expected (at 68 per cent confidence) at $z\simeq 5$ using high
resolution spectra with a total redshift path length of $\Delta z=4$
and a typical signal-to-noise ratio of $\rm S/N=15$ per
pixel. Estimates for the expected systematic uncertainties are
comparable, such that existing data should enable a measurement of
\uo\ to within $\sim 30$ per cent. This translates to distinguishing
between reionisation scenarios with similar instantaneous temperatures
at $z\simeq 5$, but with an energy deposited per proton that differs
by $2$--$3\, \rm eV$ over the redshift interval $5\la z \la 12$.  For
an initial temperature of $T\sim 10^{4}\rm\,K$ following reionisation,
this corresponds to the difference between early ($z_{\rm re}=12$) and
late ($z_{\rm re}=7$) reionisation in our models.
\end{abstract}



\section{Introduction}

The intergalactic medium (IGM) probed by the \Lya forest of absorption
lines is a valuable cosmic laboratory for studying the thermal and
ionisation history of the Universe at redshifts $z \leq
7$. Observations of intergalactic absorption lines in high redshift
quasar spectra indicate the first luminous sources had reionised the
neutral hydrogen by $5.5\leq z \leq 7$ and photo-heated the IGM to
$\sim {10}^{4}\rm\,{K}$ \citep{Becker2015}.  The recently updated
Thomson scattering optical depth reported by the {\citet{Planck_2016}
  is furthermore consistent with an instantaneous reionisation at
  $z_{\rm re}=8.8\pm0.9$. In combination with other,
  complementary observations, these observations translate to an \HI
  reionisation era that may have started as early as redshift $z\sim
  12$ and ended by $z=5.5$--$6$
  \citep{Robertson_2015ApJ,Bouwens2015,Mitra2015}.

Despite this progress, details regarding the precise timing and
duration of reionisation remain elusive.  One possible approach to
clarifying this situation is measuring the energy deposited into the
low density IGM by photo-heating during reionisation
\citep{Miralda_1994MNRAS}.  At a redshift interval $\Delta z \simeq
1$--$2$ after reionisation the temperature of the low density ($\Delta
= \rho/\bar{\rho} \leq 10$) IGM traced by the \Lya forest is expected
to follow a power law relationship, $T=T_0\Delta^{\gamma-1}$,
parameterised in terms of the temperature at the mean cosmic gas
density, $T_0$, and a slope, $\gamma-1$
\citep{Hui_1997MNRAS,McQuinnSanderbeck2016}. This temperature-density
relation has been measured using a wide variety of techniques over the
last two decades. These include analysing the velocity (Doppler)
widths of \Lya absorption lines
\citep{Haehnelt_1998MNRAS,Schaye_2000MNRAS,Ricotti_2000ApJ,McDonald2001,Rudie_2012ApJ,Bolton_2012MNRAS,Bolton_2014MNRAS},
the suppression of small-scale power in the \Lya forest flux power
spectrum
\citep{Zaldarriaga_2001ApJ,Croft_2002ApJ,Zaroubi2006,Viel_2013PhRvD},
the probability distribution of wavelet amplitudes
\citep{Meiksin2000,TheunsZaroubi2000,Zaldarriaga2002,Lidz_2010ApJ,Garzilli_2012MNRAS},
the probability distribution of the transmitted \Lya forest flux
\citep{Lidz_2006ApJ,Bolton2008,Calura2012,Lee_2015ApJ}, and the
curvature of the \Lya forest transmission
\citep{Becker_2011MNRAS,Boera2014,Boera_2016MNRAS}. The common element
to almost all these studies is that they rely on mock \Lya forest
spectra -- typically drawn from cosmological hydrodynamical
simulations -- that can be compared directly to the observational
data.

The bulk of these measurements are at redshifts $z<4$ where high
quality spectroscopic data are most readily available. These provide a
valuable probe of photo-heating during the epoch of (likely quasar
driven) \HeII reionisation around $z\simeq 3$
\citep{McQuinn2009,Compostella_2014MNRAS,Puchwein_2015MNRAS}. Importantly,
however, the long cooling timescale of the low density IGM enables
$T_0$ measurements at $z\simeq5-6$ to be used as a probe of \HI
reionisation at $z>6$
\citep{Haehnelt_1998MNRAS,Theuns_2002ApJ,Hui_2003ApJ,Trac_2008ApJ,Cen_2009ApJ,Furlanetto_2009ApJ,Lidz_2014ApJ,Daloisio_2015ApJ}. Indeed,
recent studies have demonstrated observational measurements of $T_0$
at $z=5$--$6$ are inconsistent with rapid ($\Delta z \simeq 2$) late
\HI reionisation occurring at $z \la 8$
\citep{Raskutti2012,Sanderbeck_2015}, although note this
  inference also depends on the typical spectral shape of the ionising
  sources during reionisation.

A wide range of reionisation scenarios therefore remain
consistent with these data, and their constraining power remains
relatively limited. Furthermore, the absorption features in the \Lya
forest are not only sensitive to the instantaneous thermal state of
the gas set by the Doppler broadening of the lines in velocity
space. The absorbing gas is also smoothed out in physical space by the
increased gas pressure following reionisation, leading to additional
broadening of the absorption features \citep[i.e. Jeans
  smoothing,][]{Gnedin_1998MNRAS,HuiRutledge1999,Theuns_2000MNRAS,Peeples_2010MNRAS,Kulkarni_2015ApJ,Garzilli_2015MNRAS}. The
long dynamical timescale for low density intergalactic gas
\citep[comparable to a Hubble time, e.g.][]{Schaye_2001ApJ} means the
precise degree of this pressure induced smoothing depends on the prior
thermal (and hence reionisation) history. Consequently, the degeneracy
between the Doppler broadening associated with the instantaneous gas
temperature and the uncertain degree of pressure smoothing in the low
density IGM is an important systematic for measurements of $T_0$ using
the \Lya forest. It is furthermore a nuisance parameter when
attempting to measure cosmological parameters and probe the nature of
dark matter with the \Lya forest power spectrum
\citep{McDonald2006,Zaroubi2006,Viel_2013PhRvD,PalanqueDelabrouille2015}.

Analysis of the typical coherence scale of \Lya absorption transverse
to the line of sight utilising close quasar pairs provides a promising
way to directly measuring the pressure smoothing scale at $z \simeq
2$--$3$ \citep{rorai_2013ApJ}.  However, the limited number of close
pairs currently known at higher redshift prevents this method from
being used at $z \simeq 5$, approaching the epoch of \HI
reionisation. The line of sight power spectrum of the transmitted flux
at $z \simeq 5$ -- a quantity widely studied at lower redshifts --
provides a potential alternative.  In common with other temperature
diagnostics, the power spectrum is sensitive to \emph{both} the
instantaneous temperature and the prior thermal history.  These
smoothing scales may be disentangled to some extent with high
resolution ($R\sim 40000$) spectra that probe wavenumbers $\log(k/\rm
km^{-1}\,s) \ga -1$ \citep[see e.g. Appendix D
  in][]{Puchwein_2015MNRAS}.  As the quantity of high resolution \Lya
forest data available at $z \simeq 5$ has increased in the last few
years \citep[e.g.][with 7 additional quasar spectra at $z>5.8$
    and 16 at $4.5<z<5.4$]{Becker_2015_GP}, a measurement of the
cumulative energy deposited into the IGM, and hence tighter
constraints on the thermal history during hydrogen reionisation may
now be feasible \citep[see also][]{Lidz_2014ApJ}.

In this work, we demonstrate that it is possible to constrain the
integrated thermal history at $z>5$ using the \Lya forest power
spectrum measured from data sets that are now comparable in size to
existing high resolution observational measurements.  Recent studies
have typically parameterised the integrated thermal history in \Lya
forest models as either a characteristic filtering scale, $k_{\rm F}$,
over which the gas is smoothed \citep[e.g][]{rorai_2013ApJ}, or as the
starting redshift of reionisation, $z_{\rm re}$, in optically thin
hydrodynamical simulations \citep{Viel_2013PhRvD}. The former approach
is well motivated, but in practice often treats the pressure smoothing
scale as a free parameter that is decoupled from the reionisation
history.  The latter approach is not optimal either, as the parameter
$z_{\rm re}$ does not uniquely define\footnote{For example, two
  reionisation models where $z_{\rm re}$ is identical but the spectral
  shape of the ionising sources is different will not have the same
  thermal history.}  the amount of energy deposited into the IGM as a
function of time.  In this work we propose instead that, aided by a
suitable grid of hydrodynamical models, one may instead infer the
cumulative energy per proton injected into a gas parcel during and
soon after reionisation -- a quantity which is more straightforward to
connect directly to reionisation models.

The structure of this paper is as follows. In
Section~\ref{sec:modelling}, we present an overview of the
hydrodynamical simulations used in this work and examine the typical
scales on which thermal broadening and pressure smoothing act on the
\Lya forest power spectrum at $z \simeq 5$.  In Section~\ref{sec:uo},
we examine the relationship between the gas density and \Lya forest
transmission power spectra and the cumulative energy per proton
injected into the IGM at mean density, $u_{0}$. In
Section~\ref{sec:forecast}, we forecast how well observations might
distinguish between different integrated thermal histories by
examining mock datasets within a Bayesian statistical framework via a
Markov Chain Monte Carlo (MCMC) analysis. We finally summarise our
conclusions in Section~\ref{sec:conclude}. Throughout this paper we
refer to comoving Mpc and kpc as ``cMpc'' and ``ckpc'',
respectively. A flat $\Lambda$CDM cosmology is adopted thoughout,
  with $\Omega_{\rm{m}}=0.26$, $\Omega_{\rm{\Lambda}}=0.74$,
  $\Omega_{\rm{b}}h^2=0.023$, $\sigma_8=0.80$, $h=0.72$ and
  ${n}_{\rm{s}}=0.96$.

\section{Modelling the \Lya forest at \z5} \label{sec:modelling}
\subsection{Hydrodynamical simulations}  \label{sec:sim}

\begin{table*}
\caption{The hydrodynamical simulations used in this work. All models
  have a box size of $10h^{-1}\rm\,cMpc$, $2\times512^{3}$ particles
  and a gas particle mass of $9.2\times10^4 h^{-1} M_{\sun} $. The
  columns in the table list the redshift of reionisation, $z_{\rm
    re}$, in the model, the scaling factors for the photoheating
    rates (see text for details), the logarithm of the temperature at
  mean density $\log {T}_0$, the slope of the temperature-density
  relation, $\gamma$, and the cumulative energy per proton, $u_{0}$,
  deposited into a gas parcel at mean density by $z=4.9$. The values
  of $\log T_0$ and $\gamma$ are estimated with a power-law fit to the
  volume weighted temperature-density plane.}  \centering
 \begin{tabular}{cccccccc}
  \hline
  \hline
  Model   & $z_{\rm{re}}$ & $\zeta$ & $\xi$ & $\rm{log}({T}_0^{z=4.9}/\rm K)$ & $\gamma^{z=4.9}$ & $u_{0}^{z=4.9}$ \evp & References\\
  \hline
  A15    & 9  & 0.30 & 0.00 & $3.68$ & $1.43$ & $3.1$  & Table 2,\,\citet{Becker_2011MNRAS} \\
  B15    & 9  & 0.80 & 0.00 & $3.98$ & $1.46$ & $5.9$  & \\
  C15    & 9  & 1.45 & 0.00 & $4.16$ & $1.47$ & $8.7$  &\\
  D15    & 9  & 2.20 & 0.00 & $4.28$ & $1.48$ & $11.5$ &\\
  E15    & 9  & 3.10 & 0.00 &  $4.38$ & $1.47$ & $14.5$ &\\
  F15    & 9  & 4.20 & 0.00 & $4.47$ & $1.47$ & $17.8$ &\\
  G15    & 9  & 5.30 & 0.00 &  $4.53$ & $1.48$ & $20.9$ &\\ 
  D13    & 9  & 2.20 & -0.45 & $4.28$ & $1.37$ & $11.5$ &\\
  D10    & 9  & 2.20 & -1.00 & $4.26$ & $1.08$ & $11.5$ & \\
  D07    & 9  & 2.20 & -1.60 & $4.25$ & $0.92$ & $11.5$ &\\ 
  \hline
  Tz15   & 15 & -- & -- & $3.92$ & $1.49$ & $12.4$ & Appendix B,\,\citet{Becker_2013MNRAS}\\
  Tz12   & 12 & -- & -- & $3.93$ & $1.50$ & $9.3$  &\\
  Tz9    & 9  & -- & -- & $3.92$ & $1.50$ & $5.2$  &\\
  Tz7    & 7  & -- & -- & $3.93$ & $1.47$ & $3.7$  &\\ 
  Tz9HOT & 9  & -- & --  & $4.21$ & $1.52$ & $11.3$ &\\
  \hline
  \hline
       \end{tabular}
\label{tab:simulation}
\end{table*}

\begin{figure*}
    \centering
    \begin{minipage}{.48\textwidth}
        \centering
        \includegraphics[width=\columnwidth,trim={0.0cm 0 0.5cm 0},clip]{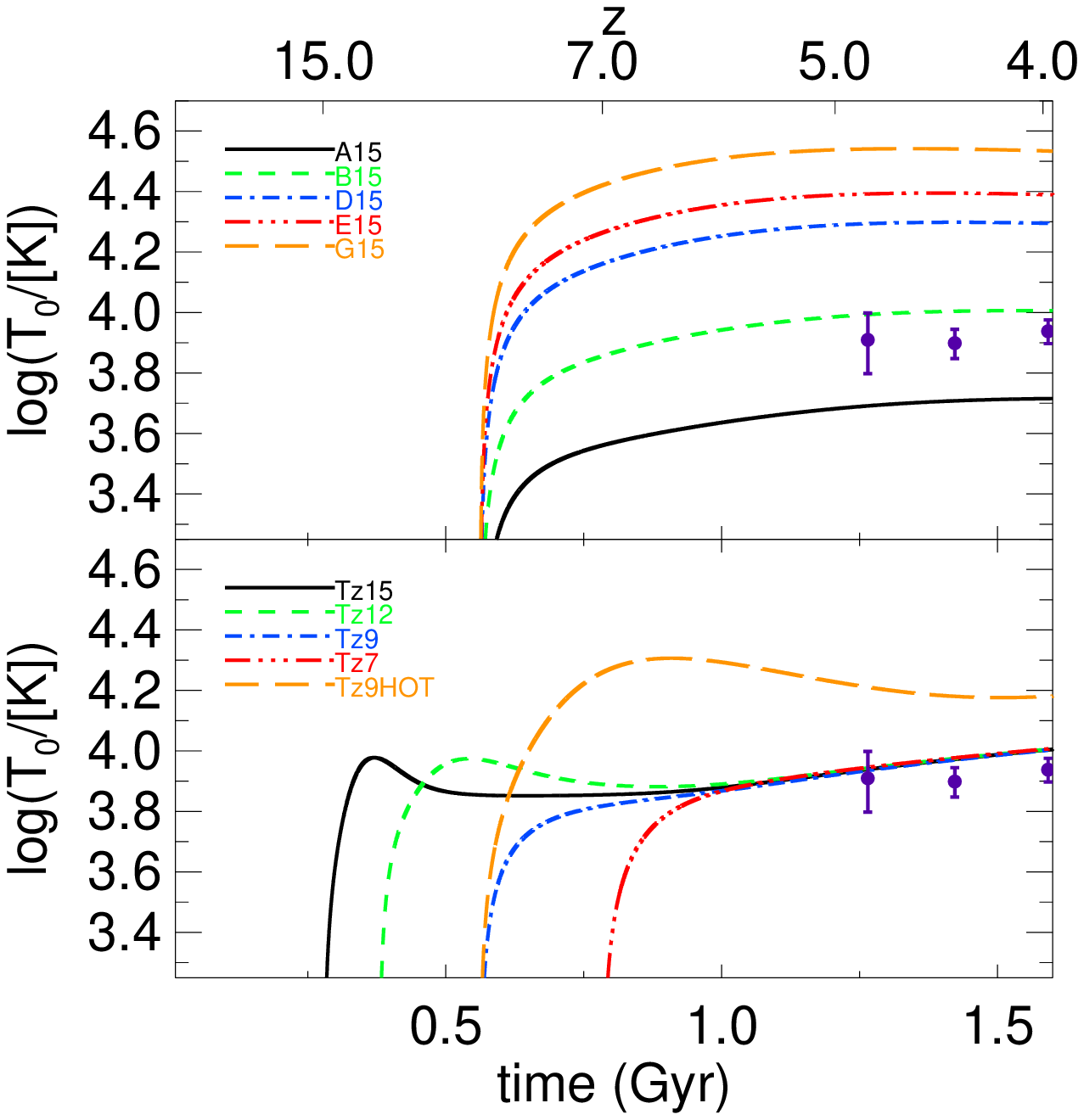}       
    \end{minipage}
    \begin{minipage}{.48\textwidth}
        \centering
        \includegraphics[width=\columnwidth,trim={0.0cm 0 0.5cm 0},clip]{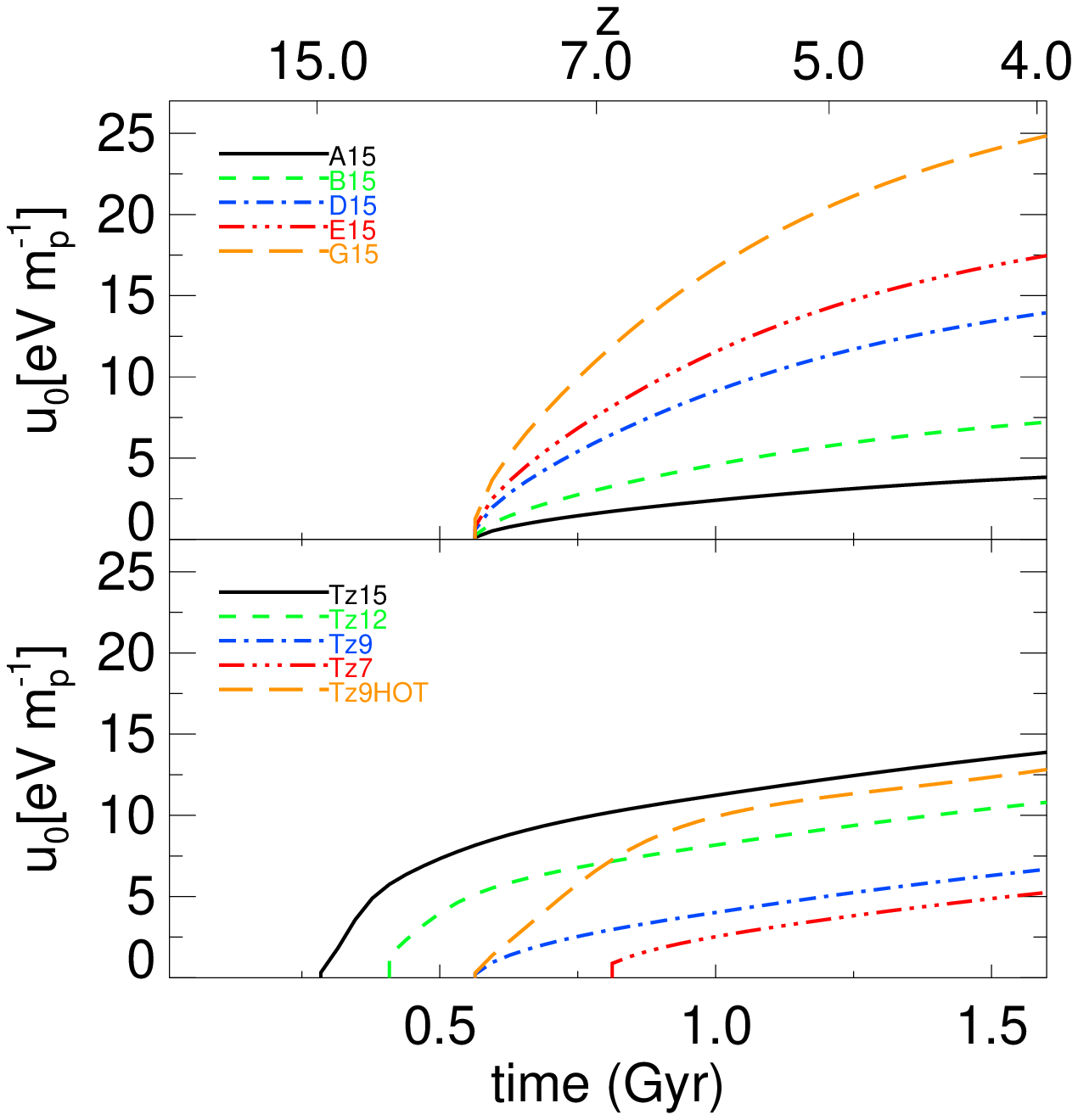}        
\end{minipage}
\vspace*{-0.5cm}
\caption{{\it Left:} The logarithm of the temperature at mean density,
  $T_0$, as function of time for a subset of the simulations listed in
  Table~\ref{tab:simulation}. The $T_0$ observational measurements
  from \citet{Becker_2011MNRAS}, evaluated at their fiducial $\gamma$
  values along with 2$\sigma$ errors, are shown by the filled
  circles. {\it Right:} The cumulative energy per proton deposited in
  a gas parcel at mean density (see Section~\ref{sec:uo} for
    details) as a function of time in the simulations.}
\label{fig:u_t_sim}
\end{figure*}

In order to model \Lya forest spectra at $z\simeq 5$ we first require
hydrodyanmical simulations with a variety of thermal histories.  The
models used in this work are summarised in Table~\ref{tab:simulation},
and are described in \citet{Becker_2011MNRAS} and
\citet{Becker_2013MNRAS}. Convergence tests with box size and mass
resolution are presented in those papers and in
\citet{Bolton_2009MNRAS}.

In brief, the simulations were performed with the smoothed-particle
hydrodynamics code \textsc{P-Gadget-3}, an updated version of the
publicly available \textsc{Gadget-2} \citep{Springel_2005MNRAS}. These
simulations use a total of $2\times512^3$ dark matter and gas
particles within a periodic $10h^{-1}\rm\,{cMpc}$ box. The initial
positions and velocities of the particles at redshift $z=99$ were
generated using the \textsc{P-Genic} initial conditions code
\citep{Springel2005nat} and the \citet{Eisenstein_1999ApJ} transfer
function.  In this work we neglect the impact of the small change in
cosmological parameters required to match the more recent results
reported by the \citet{Planck_2015}, but expect that this will not
affect our general conclusions.  The baryons in the \Lya forest
simulations are of primordial composition with a helium fraction by
mass of $Y=0.24$ \citep{OliveSkillman2004}.  Any gas particles with an
overdensity $\Delta>10^{3}$ and temperature $T<10^{5}\rm\,K$ are
converted to collisionless star particles \citep{Viel2004}.  The gas
is also photo-ionised and heated by a spatially uniform metagalactic
UV background (UVB) applied in the optically thin limit. The gas is
assumed to be in ionisation equilibrium \citep{Katz_1996ApJ} using the
recombination, ionisation and cooling rates listed in
\citet{Bolton2007nz}.

The UVB for the \citet{Becker_2011MNRAS} simulations is based on the
\citet{HaardtMadau01} synthesis model.  This includes ionising
emission from young star forming galaxies and quasars, and results in
rapid reionisation at $z_{\rm re}=9$. The photo-heating rates in most
of these models have been rescaled to reproduce a range of
temperature-density relations, such that $\epsilon_{\rm
    i}=\zeta\Delta^{\xi}\epsilon_{\rm i}^{\rm HM01}$, where
  $\epsilon_{\rm i}^{\rm HM01}$ are the \citet{HaardtMadau01}
  photoheating rates for species $i=[\rm H\,{\rm \scriptstyle I},\,
    \rm He\, {\scriptstyle I},\, He\, {\scriptstyle II}]$ and $\zeta$,
  $\xi$ are constants listed in Table~\ref{tab:simulation}.  We also
include five simulations from \citet{Becker_2013MNRAS}. These have UVB
models that have been tuned by hand to reproduce a range of
reionisation histories. Four of the models are designed to have
similar temperatures at \z5 that match the \citet{Becker_2011MNRAS}
IGM temperature measurements, but with $z_{\rm re}=[15,\,12,\,9,\,7]$.
The final model, Tz9HOT, is similar to Tz9 but with increased
photo-heating rates. The evolution of the temperature and the
cumulative energy per proton deposited in a gas parcel at the mean
background density (see Eq.~(\ref{eq:uo}) and Section~\ref{sec:uo}
  for details) in these models is displayed in
Fig.~\ref{fig:u_t_sim}.


In order to extract mock spectra from our simulations we analyse
snapshots at $z=4.915$. The spectra consist of $2048$ pixels drawn
along $1000$ random sight-lines parallel to the box boundaries. The
mean transmission, $\langle F \rangle$, of the spectra is rescaled to
correspond to an effective optical depth $\tau_{\rm eff}=-\ln\langle F
\rangle= 1.53$ \citep{Fan2006,Becker2013_taueff}, and the spectra are
convolved with a Gaussian instrumental profile with $\rm
FWHM=7\rm\,km\,s^{-1}$.  In order to aid intuition,
Fig.~\ref{fig:opt_cont} demonstrates the range of gas densities the
\Lya forest is sensitive to at $z=4.9$.  We plot the optical depth
weighted gas overdensity, ${\Delta}_{\tau}$ \citep{Schaye_1999MNRAS},
against the transmitted flux from the D15 model. The \Lya forest at
high redshift predominately probes gas close to the mean background
density, with very little contribution from regions with overdensities
greater than a few except where the transmission is saturated
($F=0$). This may be contrasted to the \Lya forest at $z=2$--$3$,
where the bulk of the transmission arises from mildly overdense gas
\citep[cf. fig. 4 in][]{Bolton_2014MNRAS}.

Finally, before proceeding further we note that one caveat to our
analysis is that reionisation is an inhomogeneous process and spatial
fluctuations in the IGM temperature and pressure smoothing scale are
expected during reionisation
\citep{Raskutti2012,Lidz_2014ApJ,Daloisio_2015ApJ}. Our
$10h^{-1}\rm\,cMpc$ simulation boxes are too small to capture this
effect -- this scale is comparable to the typical size of individual
\HII regions during reionisation
\citep[e.g.][]{Wyithe2004,Furlanetto2006} -- but for this reason
approximating a uniform redshift of reionisation over this volume is
likely reasonable.  The large scales on which temperature fluctuations
occur also translate to a modest effect ($<5$ per cent) on the one
dimensional power spectrum \citep{Lai2006,McQuinn2011,Greig2015}.  A
direct comparison of the gas clumping factor predicted by radiation
hydrodynamical simulations performed in similar volumes to this work
\citep{Finlator2012} to optically thin models \citep{Pawlik_2009MNRAS}
also yields good agreement \citep[see fig. 5
  in][]{Finlator2012}. Nevertheless, full radiation hydrodynamical
simulations that model patchy reionisation may eventually be
required. The first steps toward such large scale simulations are
being made \citep{Gnedin2014,Norman2015,Pawlik_2015MNRAS,Park2016},
although attaining the required mass resolution for modelling the high
redshift \Lya forest in large volumes remains challenging.

\begin{figure}
\includegraphics[width=\columnwidth]{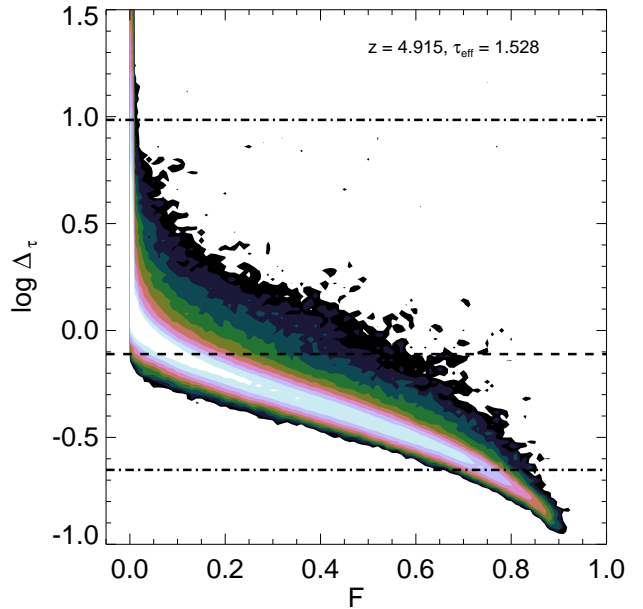}
\vspace*{-0.6cm}
\caption{Contour plot of the optical depth weighted gas overdensity,
  $\Delta_{\rm{\tau}}$, against transmitted flux $F$ at $z=4.9$ for
  the D15 model for $\tau_{\rm eff}=1.53$.  Here $\Delta_{\rm{\tau}}$
  is calculated as a weighted average ${\Delta}_{\rm{i}}=\sum
  {\tau}_{\rm{i}} {\Delta}_{\rm{i}} /\sum {\tau}_{\rm{i}}$, where
  $\Delta_{\rm{i}}$ and ${\tau}_{\rm{i}}$ are the gas overdensity and
  optical depth at the $i$th pixel on each sightline
  \citep{Schaye_1999MNRAS}.  The number density of pixels increase by
  $0.5$ dex within each contour level. The dashed and dot-dashed lines
  display the median and $95$ per cent range of the optical depth
  weighted densities.}
\label{fig:opt_cont}
\end{figure}

\subsection{The broadening of \Lya forest absorbers} 
\label{sec:broadening}

In this section we briefly review the impact of thermal broadening and
pressure smoothing on the \Lya forest power spectrum at $z \simeq 5$
\citep[see also][]
      {Bi_1992A&A,Peeples_2010MNRAS,Kulkarni_2015ApJ,Garzilli_2015MNRAS,Puchwein_2015MNRAS}. We
      begin with the assumption that \Lya absorbers are in hydrostatic
      equilibrium \citep{Schaye_2001ApJ}. The scale where the
      dynamical time equals the sound crossing timescale is the Jeans
      scale, $L_{\rm J}$, which may also be written in terms of a line
      of sight velocity, $\sigma_{\rm{J}}=H(z)L_{\rm{J}}$, where
        $L_{\rm J}$ is a proper distance.  For gas with temperature
      $T$ and normalised density $\Delta$, this corresponds to:

\begin{equation}
\begin{split}
{\sigma}_{\rm{J}} &=  \left(\frac{40 {\pi}^{2} k_{\rm{B}}}{9\mu m_{\rm{H}}}\right)^{1/2} {T}^{1/2} {\Delta}^{-1/2}  \left(\frac{{\Omega}_{\rm{m}}(1+z)^3+{\Omega}_{\rm{\Lambda}}}{{\Omega}_{\rm{m}}(1+z)^3}\right)^{1/2} \\
    &\approx 77.1 \rm{km}\,\rm{s}^{-1}\, \left(\frac{T_{0}}{{10}^4K}\right)^{1/2} \Delta^{\gamma/2-1}, \label{eq:Jeans}
\end{split}
\end{equation}

\noindent where we assume $\mu=0.61$ for the mean molecular weight of
an admixture of ionised hydrogen and singly ionised
helium\footnote{The Jeans scale in Eq.~(\ref{eq:Jeans}) is larger
    than the classical cosmological Jeans scale, $\lambda_{\rm J}$ --
    derived from linear theory when assuming an adiabatic thermal
    history -- by a factor of $2\pi$
    \citep{Bi_1992A&A,Kulkarni_2015ApJ}. For arbitrary thermal
    histories within the linear theory derivation,
    \citet{Gnedin_1998MNRAS} further show that the pressure smoothing
    may be described by a filtering scale, $\lambda_{\rm F}$, which
    depends on the prior thermal history.  Typically $\lambda_{\rm
      F}<\lambda_{\rm J}$ and $\lambda_{\rm F}\sim 100 \rm\, ckpc$
    ($\sim 10\rm\,km\,s^{-1}$ at $z=5$), although the precise value is
    dependent on the prior heating history of the IGM.}.  In the
second line we have also used the fact that $T=T_{0}\Delta^{\gamma-1}$
and $\Omega_{\rm m}(1+z)^{3} \gg \Omega_{\rm \Lambda}$ at $z \ga
3$. Note, however, the Jeans scale only approximates the pressure
smoothing scale in the low density IGM.  As the dynamical timescale,
$t_{\rm dyn} = \sqrt{\pi/G\rho_{\rm m}} \simeq
H(z)^{-1}\Delta^{-1/2}$, is long for low density gas the absorbing
structures in the \Lya forest at $z \simeq 5$ will not have reached
hydrostatic equilibrium. The pressure smoothing scale is instead
better described as $\sigma_{\rm p}=f_{\rm J}\sigma_{\rm J}$, where
$f_{\rm J}<1$ and depends on the prior thermal history
\citep[][and see footnote 2]{Gnedin_1998MNRAS,HuiRutledge1999}.


\comment{\begin{equation}
\frac{1}{k^2_{\rm F}} =  \frac{4\pi G\bar{\rho_{m,0}}}{b(t)} \int_0^t dt^{\prime} \frac{b(t^{\prime})}{a(t^{\prime})k^2_{\rm J}(t^{\prime})} \int_{t^{\prime}}^{t^{\prime \prime}} \frac{dt^{\prime \prime}}{a^2(t^{\prime \prime})} 
\end{equation}

\noindent here, $\rho_{m,0}$ is matter density at $z=0$ and $b(t)$ linear perturbation growth factor. }

In comparison, the thermal (or Doppler) broadening scale for a
Gaussian line profile is given by:

\begin{equation}
  \sigma_{\rm{th}} = \left(\frac{k_{\rm{B}}T}{m_{\rm{H}}}\right)^{1/2}=9.1\rm{km}\,\rm{s}^{-1}\,\left(\frac{T_{0}}{{10}^4\,\rm K}\right)^{1/2}\Delta^{(\gamma-1)/2}. 
\end{equation}

\noindent
The ratio of these two scales is $\sigma_{\rm p}/\sigma_{\rm th}\simeq
8.5 f_{\rm J}\Delta^{-1/2}$. In general we therefore expect the
pressure smoothing to act on similar scales to the thermal
  broadening.  Fortunately, as we see shall see next, the different
  scale dependence of these effects in our hydrodynamical simulations
  at $z=5$ enables us to break this degeneracy.

\subsection{The line of sight \Lya forest power spectrum}\label{sec:pkplot}

We compute the power spectrum of the transmitted flux,
$P_{\rm{F}}(k)$, at $z=4.9$ from our simulations using the estimator
$\delta_{\rm F}=F/<F>-1$, where $\langle F\rangle =\langle e^{-\tau}
\rangle$ is the mean transmission (or equivalently the effective
optical depth, $\tau_{\rm eff}=-\ln \langle F \rangle=1.53$) of the
$1000$ sight-lines drawn from each simulation.  The top row of
Fig.~\ref{fig:powerspec} shows the results for a sub-set of the models
listed in Table~\ref{tab:simulation}. The left hand panel displays the
effect of changing $T_0$ on the power spectrum; higher temperatures
result in decreased power at wavenumbers $\log(k/\rm km^{-1}\,s)>-1.5$
arising from a combination of thermal broadening and pressure
smoothing.  The middle panel demonstrates the effect of changing
$\gamma$ -- the slope of the temperature-density relation -- is more
modest, with a slight increase in power over all scales as $\gamma$ is
decreased.  This is in part due to the fact that the typical gas
densities probed by the \Lya forest at $z\simeq 5$ are close to mean
density, and the characteristic pressure and thermal broadening scales
both have a modest dependence on gas density.  It also suggests that
any constraint on $\gamma$ from $P_{\rm{F}}(k)$ is likely to be weak
at this redshift.

\begin{figure*}
\includegraphics[trim=0.0cm 0.0cm 3.5cm 1.2cm, clip=true, width=0.95\textwidth]{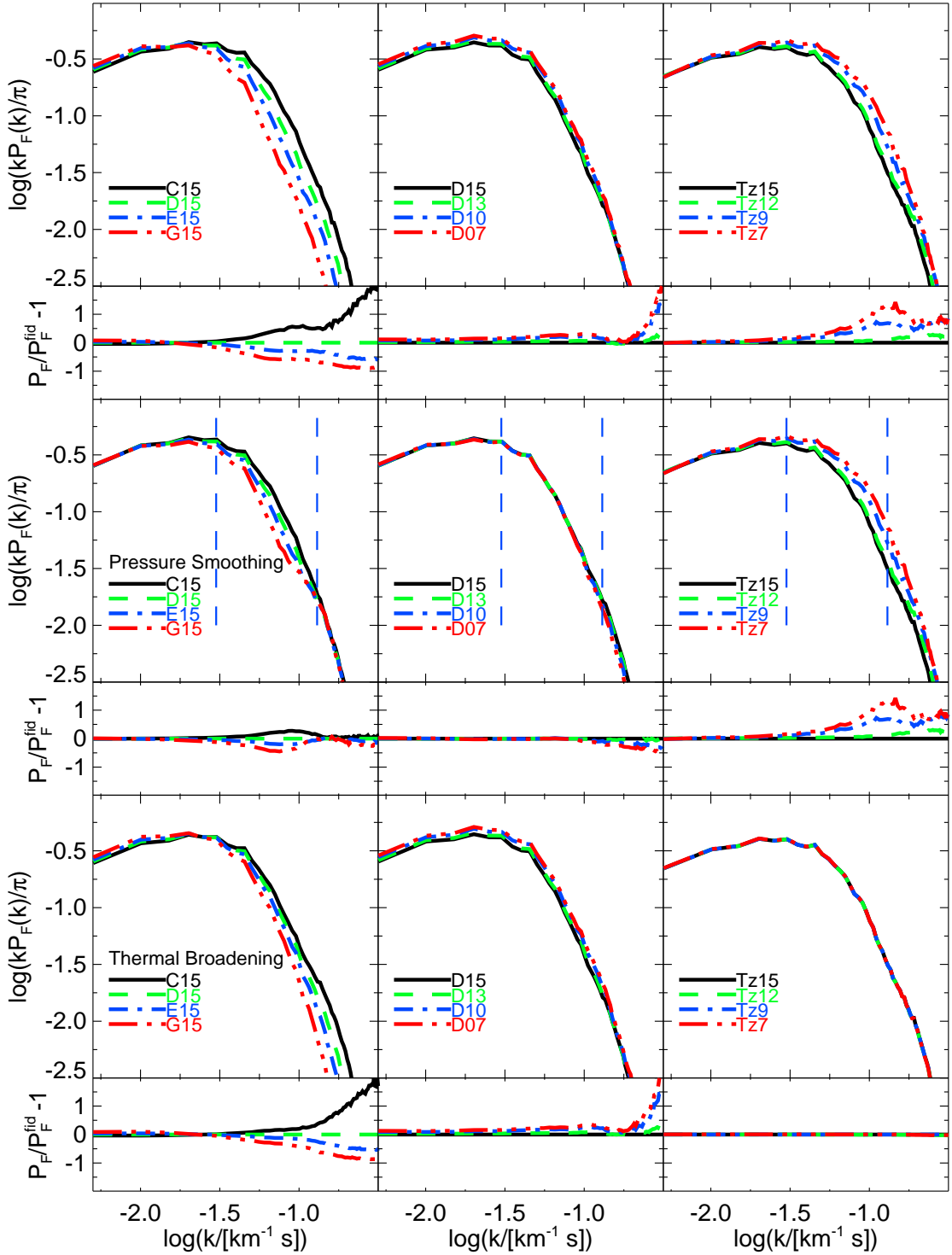}
\vspace*{-5mm}

\caption{{\it Top row:} the transmitted flux power spectrum --
  including variations from both pressure smoothing and thermal
  broadening -- at $z=4.9$ for a sub-set of models with varying
  $T_{0}$ (left), $\gamma$ (middle) and the redshift of reionisation
  (right).  The power spectra are displayed relative to the D15 (left
  and middle) and Tz15 (right) models are displayed immediately
  below. {\it Middle row:} the power spectrum for the same
  simulations, but now with each $T-\Delta$ relation mapped to the D15
  (left and middle) and the Tz15 model (right). The thermal broadening
  in these models is therefore identical. The dashed blue lines
  display the approximate wavenumber range over which pressure
  smoothing is dominant. Note that for simulations with varying
  $\gamma$ (D15-D07, middle column), pressure smoothing has very
  little effect on the power spectrum except at the smallest
  scales. {\it Bottom row:} the transmitted flux power spectrum for
  the D15 (left and middle) and Tz15 (right) models after imposing the
  $T-\Delta$ relation from the models indicated in the figure legend.
  The pressure smoothing in these models is identical. The varying
  $z_{\rm re}$ models have almost identical values of $T_0$ at $z \sim
  4.9$ (Fig.~\ref{fig:u_t_sim}) and therefore are indistinguishable
  when pressure smoothing is removed. This can be seen by comparing
  middle-right and bottom-right panels.  The \uo, $T_{0}$ and
    $\gamma$ values for each model are listed in
    Table~\ref{tab:simulation}. All mock spectra are scaled to have
  $\tau_{\rm eff}=1.53$, and have been convolved with a Gaussian with
  FWHM=$7\rm\, {km\,s}^{-1}$.}
\label{fig:powerspec}
\end{figure*}

The right panel in the top row displays the four models with varying
$z_{\rm re}$; recall these have similar $T_0$ at $z=4.9$ but different
reionisation redshifts.  Any differences in $P_{\rm F}(k)$ are due
variations in the pressure smoothing scale only. The Tz15 model has
less power (and more pressure smoothing) than the Tz7 and Tz9 models
over a wide range of wavenumbers, with the largest differences at
$\log(k/\rm km^{-1}\,s)\simeq -1$. Earlier reionisation allows more
time for the gas to respond to the change in pressure due to heating
during and soon after reionisation, resulting in increased smoothing
of the gas distribution. Note also the power spectra for the Tz15 and
Tz12 models are very similar, although the cumulative energy per
proton deposited at mean density, $u_{0}$, by $z=4.9$ in these models
is rather different.  A related result was noted by
\citet{Pawlik_2009MNRAS}, who found that the clumping
factor\footnote{The clumping factor is related to the root mean square
  of the density contrast (and hence also the gas density power
  spectrum) by $\langle \delta^{2} \rangle=C-1$, where the density
  contrast is $\delta = \rho/\bar{\rho}-1$.}, $C=\langle \rho^{2}
\rangle / \bar{\rho}^{2}$, of gas in optically thin hydrodynamical
simulations at $z\approx6$ is insensitive to the redshift of
reionisation if $z_{\rm re} \geq 9$.  Although the exact upper
redshift limit will be model dependent, this indicates the pressure
smoothing is only sensitive to the prior IGM thermal history over a
limited redshift range (see also Figure~\ref{fig:powerspecz} and text
in Section~\ref{sec:conclude}).

\begin{figure*}
    \centering
\begin{minipage}{0.33\textwidth}
        \centering
        \includegraphics[width=\columnwidth,trim=0.5cm 0.0cm 0.3cm 1.1cm, clip=true]{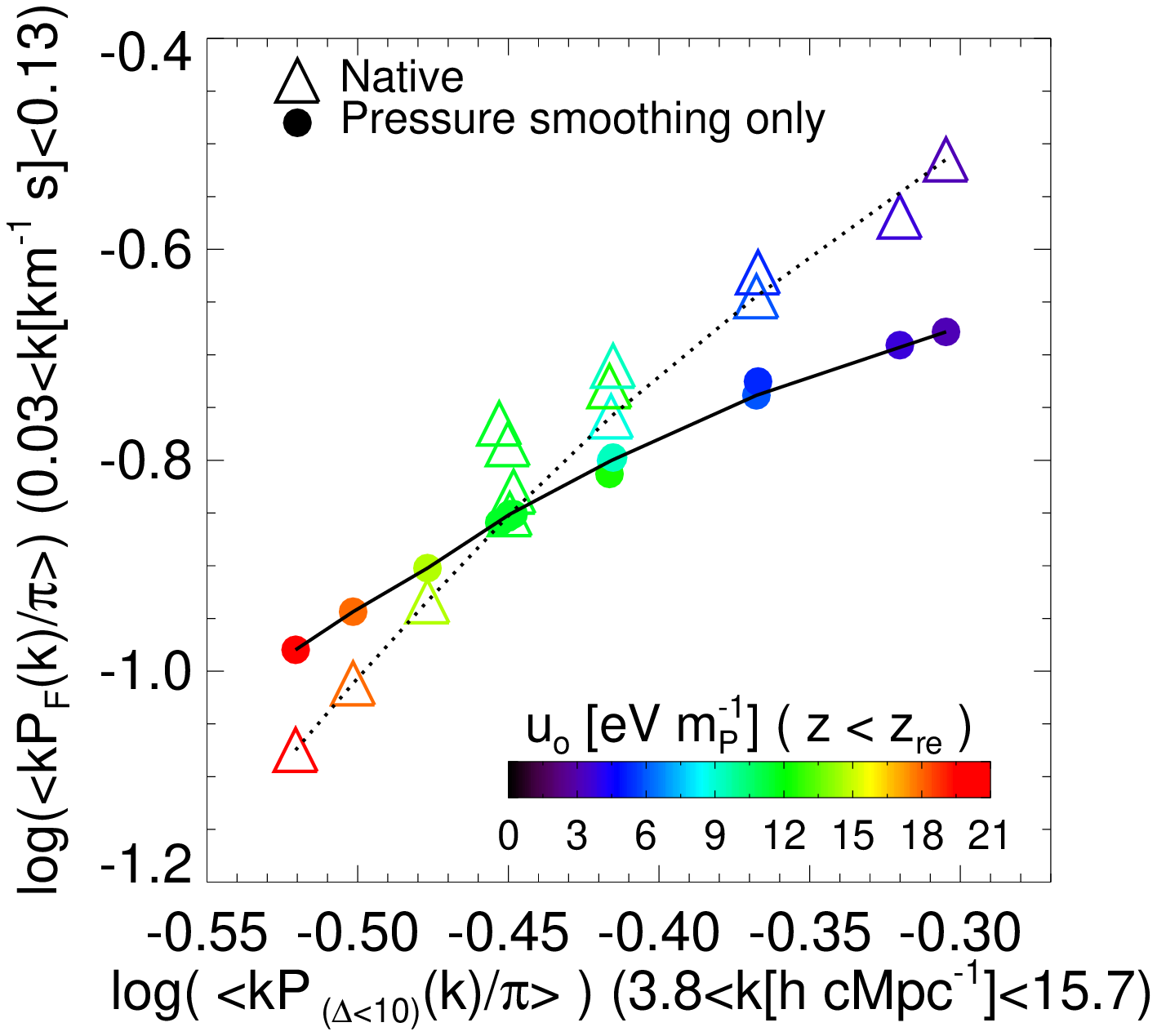}       
\end{minipage}
\begin{minipage}{0.32\textwidth}
        \centering
        \includegraphics[width=\columnwidth,trim=1.0cm 0.0cm 0.5cm 1.0cm, clip=true]{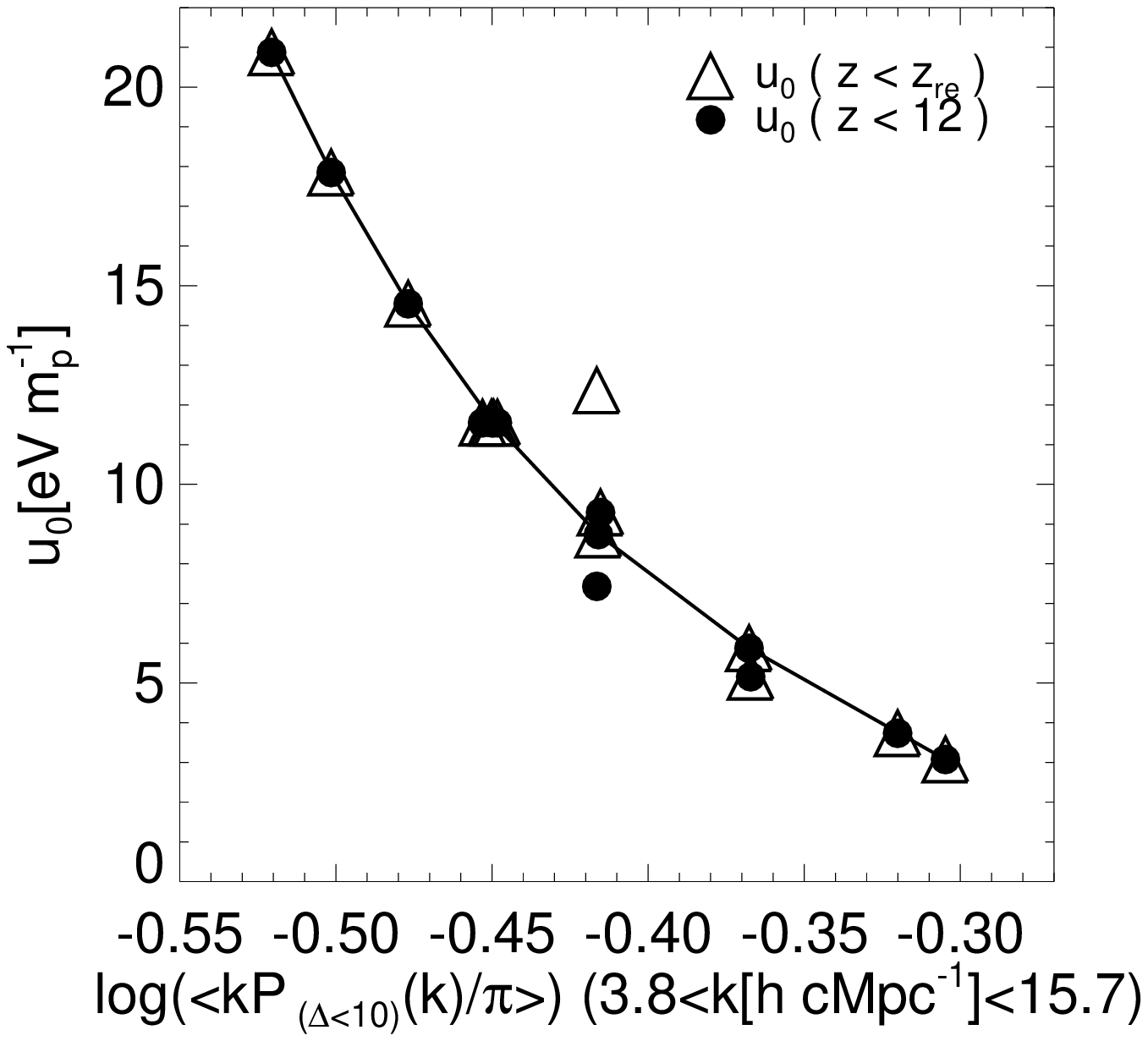}        
\end{minipage}
\begin{minipage}{0.32\textwidth}
        \centering
        \includegraphics[width=\columnwidth,trim=0.5cm 0.0cm 1.4cm 1.0cm, clip=true]{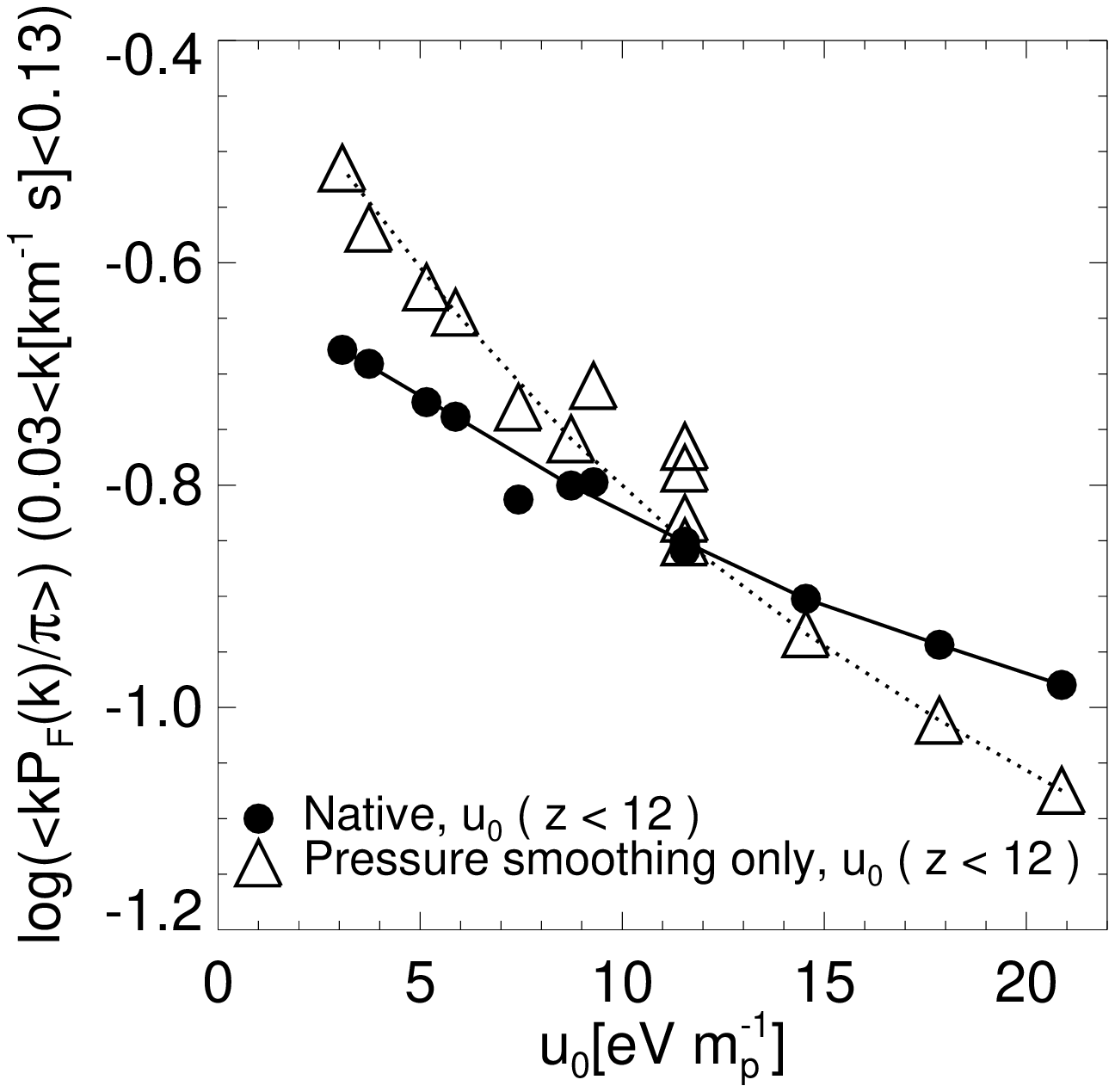}        
\end{minipage}
\vspace*{-0.5cm}
\caption{{\it Left:} The open triangles display the average of the
  \Lya forest power spectrum at wavenumbers $0.03<\mathrm{k}/\rm
  km^{-1}\,s <0.13$ (approximately the scale where pressure smoothing
  dominates) against the $\Delta<10$ gas density power spectrum
  averaged over the equivalent scale.  The A15--G15, Tz15--Tz7 and
  D13--D07 models are shown at $z=4.9$, with a dotted curve through
  the A15--G15 models.  The average \Lya forest power spectrum for
  each model after thermal broadening differences are removed is shown
  by the filled circles. Here each $T-\Delta$ distribution is mapped
  to the $T_0$ and $\gamma$ for the D15 model, and a solid curve is
  drawn through the A15--G15 models.  The colour scale indicates \uo
  for each model. {\it Centre:} The cumulative energy per proton
  against the average of the $\Delta<10$ density power spectrum
  computed using Eq.~(\ref{eq:uo}) for all photo-heating up to \zre
  for each model (open triangles) and $z=12$ (filled circles). The
  solid curve is again drawn through the A15--G15 models. {\it
      Right:} The average of the \Lya forest flux power spectrum at
    wavenumbers $0.03<\mathrm{k}/\rm km^{-1}\,s <0.13$ (open
    triangles) against $u_{0}(z<12)$.  The filled circles show the
    same quantity once differences due to thermal broadening are
    removed.  The solid and dashed curves are again drawn through
    A15--G15 models. }
\label{fig:flux_den_uo} 
\end{figure*}

We may examine the impact of pressure smoothing and thermal broadening
on the \Lya forest power spectrum more easily by separating these
effects in our models.  We first fit a single power law to the
$T-\Delta$ relation in each model, with \logto\ and $\gamma-1$ as the
intercept and slope.  We then translate and rotate the entire
$T-\Delta$ plane in each simulation to match \logto\ and $\gamma-1$
from another model.  This procedure allows us to change the
instantaneous temperature of the gas, but retain the same pressure
smoothing scale (which arises from the underlying gas density
distribution). The middle row in Fig.~\ref{fig:powerspec} displays the
result of this procedure, where we have transformed each $T-\Delta$
plane in each model to correspond to the $T_0$ and $\gamma$ values in
the D15 simulation in the left and middle column, and the Tz15 model
in the right.  Note that as the temperatures are changed we also
rescale the neutral hydrogen number densities in the simulated spectra
as $n_{\rm{HI}}\propto T^{-0.72}$, due to the temperature dependence
of the \HII recombination coefficient \citep{VernerFerland1996}.  All
models are again rescaled to have the same $\tau_{\rm eff}=1.53$.

As might be expected, the different thermal histories in the
simulations displayed in the middle left panel of
Fig.~\ref{fig:powerspec} produce rather different pressure smoothing
scales.  With the effect of thermal broadening removed, this effect is
most prominent at wavenumbers $0.03 \leq k/[\rm km^{-1}\,s] \leq
0.13$, shown by the dashed vertical lines, although it operates to a
lesser extent at smaller scales (i.e. larger wavenumbers) as well.  In
contrast, the central panel demonstrates the slope of the $T-\Delta$
relation has very little impact on the pressure smoothing except at
the smallest scales -- note the cumulative energy per proton deposited
into a gas parcel at mean density is identical in these
simulations. The models with varying $z_{\rm re}$ are also largely
unchanged, emphasizing again that it is the pressure smoothing which
causes the differences in the power spectrum for these models.

Finally, the bottom row of Fig.~\ref{fig:powerspec} displays the flux
power spectrum computed using the density field from the D15 model
(left and middle panel) and the Tz15 model (right panel), but with an
imposed $T-\Delta$ relation that matches the models indicated in the
figure panels. This procedure isolates the impact of thermal
broadening on $P_{\rm F}(k)$.  There is some degeneracy with the
pressure smoothing, but in general the thermal broadening acts on
smaller scales, with the largest difference in the models occurring at
$\log(k/\rm km^{-1}\,s) > -1$.  The small-scale cut-off for the power
spectrum is mainly determined by the instantaneous temperature
\citep{Peeples_2010MNRAS}.  This also suggests that measurements of
the power spectrum at small scales, $-1 \leq \log(k/\rm km^{-1}\,s)
\leq -0.5$, are required to break the degeneracy between pressure
smoothing and thermal broadening.  Note also the models in the middle
panel are similar to the results in the top row; most of the
contribution to the power when changing $\gamma$ is from thermal
broadening.  As expected, there is no apparent difference in power
among the varying $z_{\rm re}$ models, which are designed to reach a
similar temperature at mean density around \z5.

\section{From flux power spectrum to thermal history}\label{sec:uo}

We now proceed to examine the relationship between the transmitted
flux power spectrum at $z\simeq 5$ and the integrated thermal history
in our hydrodynamical simulations.  The temperature evolution of a gas
parcel with density $\rho$ in an expanding universe can be expressed
as \citep[e.g.][]{Miralda_1994MNRAS,McQuinnSanderbeck2016}

\begin{equation}
 \frac{dT}{dt} = \frac{2\mu m_H}{3k_B\rho}(\mathscr{H}-\Lambda)+\frac{2T}{3(1+\delta)}\frac{d\delta}{dt}+\frac{T}{\mu}\frac{d\mu}{dt}-2HT,
\end{equation}
\label{eq:rtc}

\noindent
where $\mathscr{H}=\sum_{\rm i}n_{i}\epsilon_{i}$ is the total
photoheating rate per unit volume for the species $i=[$H$\,\rm
  \scriptstyle I,\,$He$\,\rm \scriptstyle I,\,$He$\,\rm \scriptstyle
  II]$, $\Lambda$ is the cooling rate per unit volume, and $H$ is the
Hubble parameter.  The first term in Eq.~(\ref{eq:rtc})
encapsulates all the photo-heating and radiative cooling
processes. The second term describes adiabatic heating and cooling
from structure formation, and the third term is associated with
changes in the mean molecular weight.  The final term arises from
adiabatic cooling due to the expansion of the Universe.




The cumulative energy deposited into a gas parcel by photo-heating is
obtained by considering the first term in Eq.~(\ref{eq:rtc}) and
setting the radiative cooling term to zero. Noting that the specific
internal energy is given by $u={3 k_B T}/{2 \mu m_{\rm{H}}}$, we may
then write $du/dt=\mathscr{H}/{\rho}$.  For a gas parcel at the mean
background density, the total energy per unit mass deposited into the
gas parcel by redshift $z_{0}$ is

\begin{equation} \label{eq:uo}
u_{0} = \int_{z_{\rm{o}}}^{z_{\rm{re}}} \frac{\mathscr{H}}{\bar{\rho}}\frac{dz}{H(z)(1+z)},
\end{equation}

\noindent
where $\bar{\rho}={\rho}_{\rm{crit}}{\Omega}_{\rm{b}}{(1+z)}^{3}$ is
the mean background baryon density.  This quantity is displayed in the
left panel of Fig.~\ref{fig:u_t_sim} and is listed in
Table~\ref{tab:simulation} at $z=4.9$.  The cumulative energy per
proton deposited into a gas parcel at mean density is straightforward
to compute for a given reionisation history in our \Lya forest
simulations.

We illustrate the relationship between the transmitted flux power
spectrum, the gas density power spectrum and $u_0$ in our
hydrodynamical simulations in Fig.~\ref{fig:flux_den_uo}.  The open
triangles in the left panel display the mean of the transmitted flux
power spectrum against the mean of the gas density power spectrum for
all gas with $\Delta<10$.  The mean is obtained over the scales $0.03
\leq k/[\rm km^{-1}\,s] \leq 0.13$, approximately corresponding to the
scales over which the influence of pressure smoothing is largest in
our models (see Fig.~\ref{fig:powerspec}).  Following
\citet{Kulkarni_2015ApJ} and \citet{Lukic2015}, we consider the gas
density power spectrum for normalised densities $\Delta < 10$ only;
including higher density gas associated with non-linear structure
results in significantly more power toward small scales.  As shown in
Fig.~\ref{fig:opt_cont}, the \Lya forest power spectrum at $z \simeq
5$ is insensitive to absorption from gas at these densities.  The
precise choice of cut-off here is somewhat arbitrary, but is motivated
by the fact that optical depth weighted densities, $0.2 \leq
\Delta_{\tau} \leq 10$, bound 95 per cent of all \Lya forest pixels at
$z=4.9$ in our models.

There is a correlation between the \Lya forest power spectrum and the
underlying gas density power spectrum, as expected.  Models with a
greater energy deposited per proton exhibit less power on scales $0.03
\leq k/(\rm km^{-1}\,s) \leq 0.13$ due to the smoother distribution of
gas.  The points that scatter upward from the dotted curve correspond
to the varying $\gamma$ and $z_{\rm re}$ models.  The increased power
in the transmitted flux arises from differences in the thermal
broadening, even for models where the average gas density power
spectrum (and energy input per proton) are similar.  The pressure
smoothing is thus still somewhat degenerate with thermal broadening on
these scales.  This is evident from the filled circles in the left
panel of Fig.~\ref{fig:flux_den_uo}, which display the average \Lya
forest power spectrum after rescaling the $T$--$\Delta$ relation in
all models to match the D15 simulation.  This implies if the
degeneracy between thermal broadening and pressure smoothing is broken
with the transmitted flux power spectrum on scales $\log(k/\rm
km^{-1}\,s) > -1$, the \Lya forest directly probes the underlying gas
density power spectrum (or equivalently the gas clumping
factor\footnote{We have verified that the gas clumping factor,
  $C=\langle \rho^{2} \rangle / \bar{\rho}^{2}$, for gas with
  $\Delta<10$ in the simulations is also tightly correlated with the
  gas density power spectrum averaged over the scales used in
  Fig.~\ref{fig:flux_den_uo}.  The clumping factor is $C \simeq
  2$--$3$ in our models at $z=4.9$.}) at $z\simeq 5$.

The open triangles in the centre panel of
Fig.~\ref{fig:flux_den_uo} display the cumulative energy deposited per
proton at mean density, $u_0$, computed using Eq.~(\ref{eq:uo})
against the gas density power spectrum for $\Delta<10$.  The gas
density power spectrum is averaged over the same scale as in the left
panel.  Again, there is an excellent correlation between the two
quantities aside from the triangle at $u_0=12.4\rm \, eV\,m_{\rm
    p}^{-1}$ corresponding to the Tz15 model with $z_{\rm re}=15$.
All the other models experience rapid reionisation at $z \leq 12$. As
discussed earlier, this is because the thermal history at $z>12$ does
not significantly impact on the pressure smoothing scale of the gas in
our simulations. This is illustrated by the filled circles in the
right panel, which show $u_{0}$ computed at $z \leq 12$ only.

Finally, the open triangles in the right panel of
  Fig.~\ref{fig:flux_den_uo} display the correlation between the
  average flux power spectrum on scales $0.03 \leq k/(\rm km^{-1}\,s)
  \leq 0.13$ and $u_0$ at $z<12$.  Note again there is some degeneracy
  with thermal broadening when averaging on these scales; the filled
  circles show the same quantity once differences due to thermal
  broadening are removed.  This simple analysis suggests that
$u_{0}(z \la 12)$ should serve as a convenient and useful
parameterisation for the prior thermal history in our
simulations.  A more rigorous approach requires analysing the full
\Lya forest power spectrum and correctly dealing with the parameter
degeneracies in the model, which we turn to next.

\section{Inferring the thermal history during reionisation}\label{sec:forecast}

\subsection{Markov Chain Monte Carlo analysis}\label{sec:mcmc}
\begin{figure*}
    \centering
    \begin{minipage}{.7\textwidth}
        \centering
        \includegraphics[trim={1.0cm 0.0cm 1.0cm 2.9cm}, clip=true, width=\columnwidth]{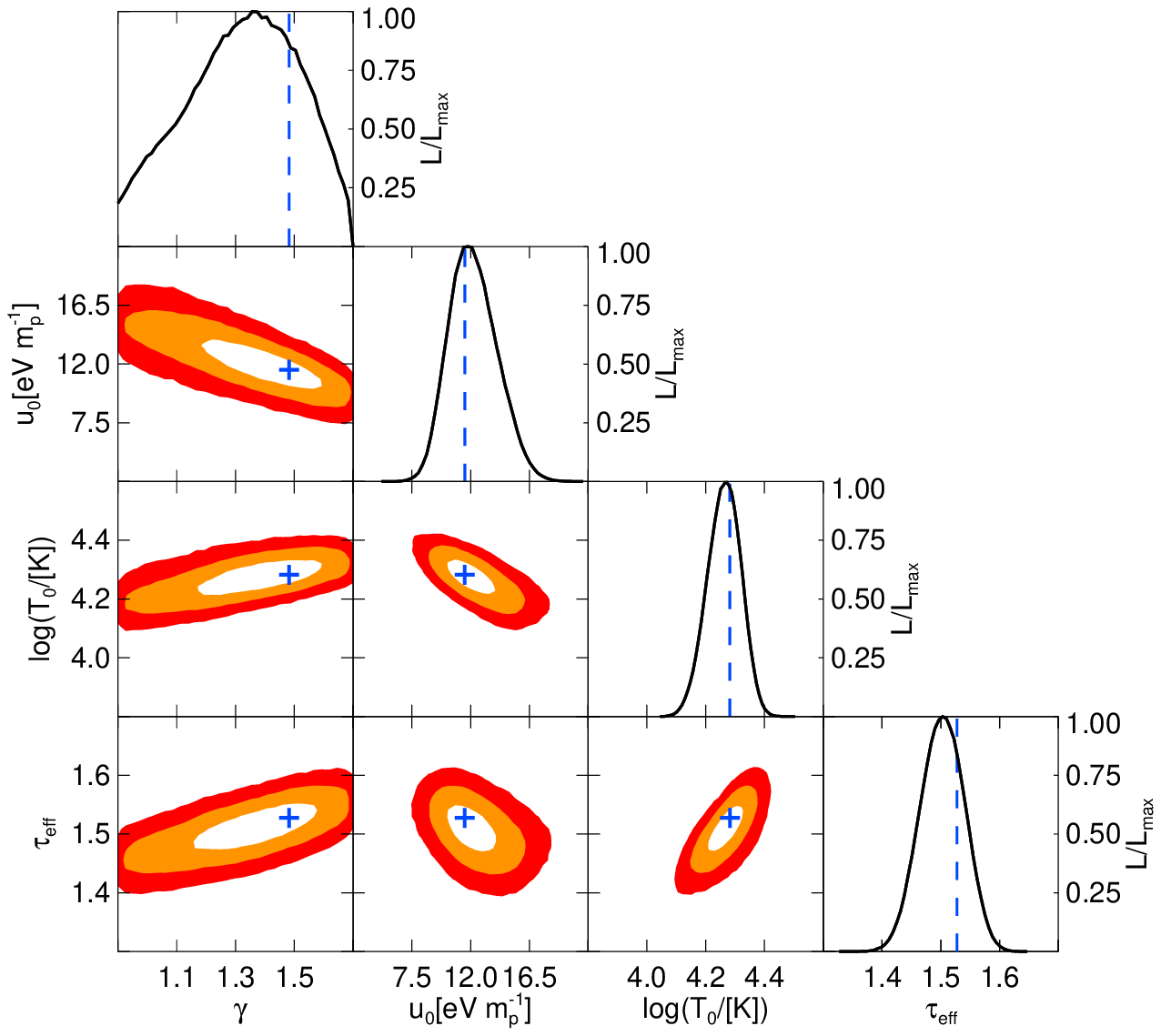}       
    \end{minipage}
    \begin{minipage}{.7\textwidth}
        \centering
        \includegraphics[trim={1.0cm 0.0cm 1.0cm 2.9cm}, clip=true, width=\columnwidth]{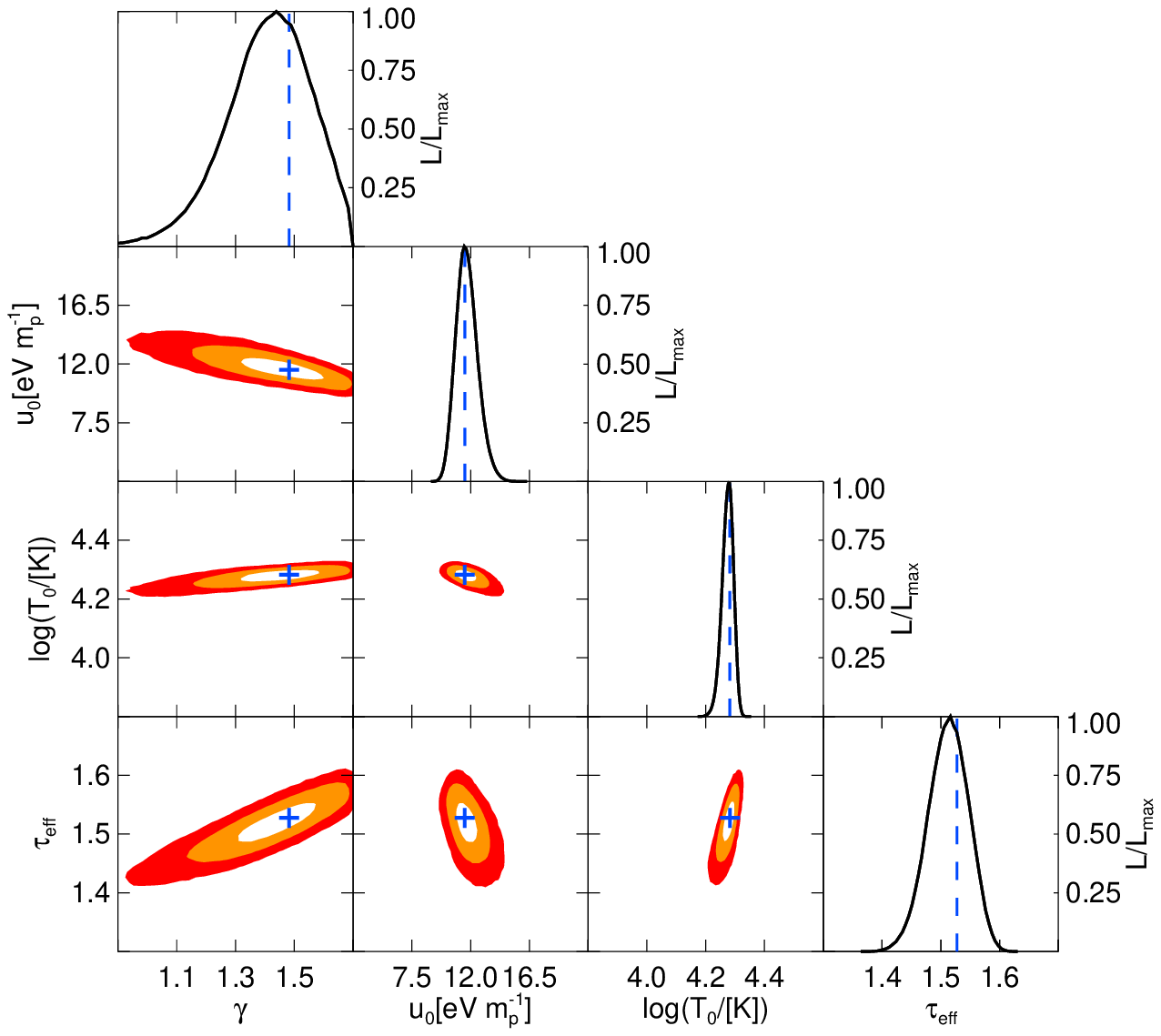}        
\end{minipage}
\vspace{-0.65cm}
\caption{{\it Top:} The contours display the two dimensional
  probability distributions for the parameters $\log T_{0}$, $u_0$,
  $\gamma$ and \taueff\ recovered from mock observations of the D15
  \Lya forest power spectrum using the realistic data scenario. The
  joint 1$\sigma$, 2$\sigma$ and 3$\sigma$ contours are shown in
  white, orange and red, respectively.  The black curves display the
  one dimensional marginalised posterior distributions for each
  parameter. The blue cross and blue vertical dashed line show the
  true model values (see Table~\ref{tab:mcmc}). {\it Bottom:} As for
  the top panel, except now for the optimistic data scenario (see text
  for further details).}
\label{fig:cpD15}
\end{figure*}

\begin{figure}
\includegraphics[width=\columnwidth,trim=1.0cm 1.0cm 2.5cm 2.0cm, clip=true,]{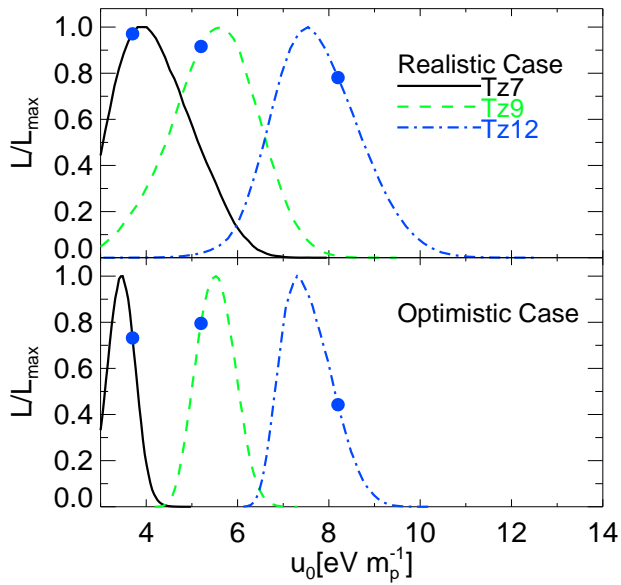}
\vspace{-1.0cm}
\caption{The one dimensional marginalised posterior distributions for
  $u_0$ obtained from mock observations of the Tz12 (solid black
  curve), Tz9 (dashed green curve) and Tz7 (dot-dashed blue curve)
  simulations. The upper (lower) panels display the realistic
  (optimistic) data scenario.  The true $u_0$ values at $z<11.5$ in
  the simulations are shown by the blue points.  These models
    have very similar values for $\rm log(T_0)$ and $\gamma$ at
    $z\simeq5$ (see Table~\ref{tab:simulation}).}
\label{fig:prob_uo} 
\end{figure}

\begin{figure*}
    \centering
    \begin{minipage}{.7\textwidth}
        \centering
        \includegraphics[trim={1.0cm 0.0cm 1.0cm 2.9cm}, clip=true, width=\columnwidth]{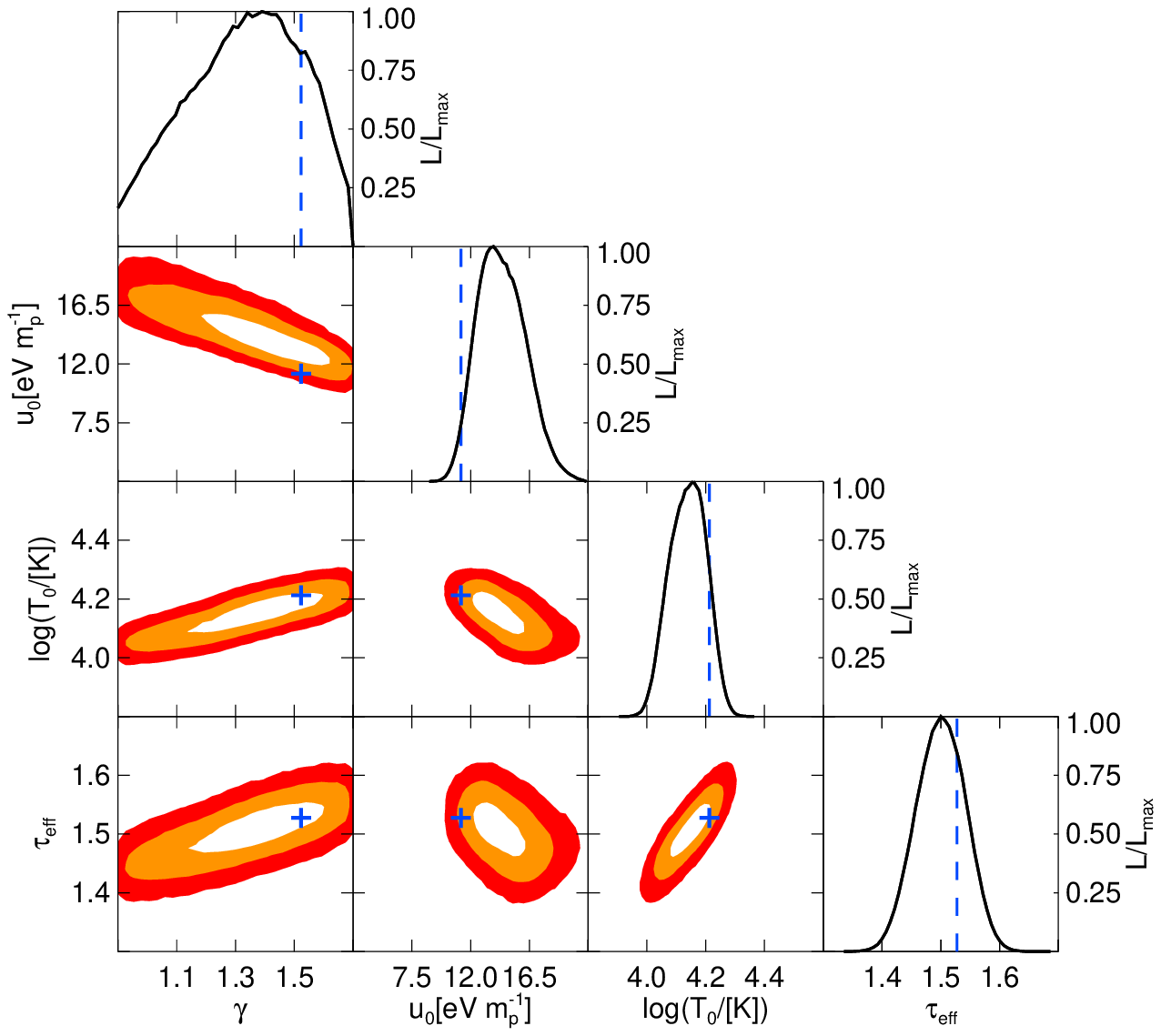}       
    \end{minipage}    
\vspace{-0.5cm}
\caption{As for Fig.~\ref{fig:cpD15}, but now for the Tz9HOT
    model using the realistic data scenario. }
\label{fig:cpTz9HOT}
\end{figure*}

We make forecasts for the constraints attainable on the thermal
history using a Bayesian MCMC approach.  Given a set of power spectrum
measurements, $\emph{P}_{\rm{F}}^{\rm{data}}$, we maximise the
likelihood function, $\pazocal{L}$, with respect to the model
parameters used in our hydrodynamical simulations, $M$,
\citep[e.g.][]{Zaroubi2006,Viel_2009,rorai_2013ApJ}

\begin{equation}
\ln \pazocal{L}({\emph{P}_{\rm{F}}^{\rm{data}}}|M) \propto (\emph{P}_{\rm{F}}^{\rm{data}} - \emph{P}_{\rm{F}}^{\rm{model}})^{T} {\Sigma}_{\rm{data}}^{-1}(\emph{P}_{\rm{F}}^{\rm{data}} - \emph{P}_{\rm{F}}^{\rm{model}}).    \label{eq:likeli}
\end{equation}

\noindent
Here $\emph{P}_{\rm{F}}^{\rm{model}}$ is the simulated \Lya forest
power spectrum for a given set of model parameters $M$, while
$\Sigma_{\rm{data}}$ is the covariance matrix for the measured power
spectrum.

We consider four parameters in our analysis -- \logto, \uo, $\gamma$
and \taueff\ -- and vary these to construct grid of models based on
our A15--G15 simulations.  We obtain combinations of the three thermal
parameters by imposing different $T$--$\Delta$ relations on the
simulations, as described in Section~\ref{sec:pkplot}.  In this way we
retain the gas density power spectrum associated with a given value of
\uo\ in our models while varying the instantaneous temperature.  We
consider seven values for for the cumulative energy deposited per
proton over the range $u_{0}=3.1$--$20.9 \rm \,eV\,m_{\rm p}^{-1}$,
following the parameter range covered by our hydrodynamical
simulations\footnote{For reference, the UVB synthesis models from
  \citet{FaucherGiguere_2009_uvb}, \citet{HaardtMadau01} and
  \citet{Haardt_2012ApJ} correspond to reionisation at $z_{\rm
    re}^{\rm FG09}=10$, $z_{\rm re}^{\rm HM01}=9$ and $z_{\rm re}^{\rm
    HM12}=15$ with $u_{0}^{\rm FG09}=7.5\, \rm eV\,m_{\rm p}^{-1}$,
  $u_{0}^{\rm HM01}=6.7\, \rm eV\,m_{\rm p}^{-1}$ and $u_{0}^{\rm
    HM12}=11.0\, \rm eV\,m_{\rm p}^{-1}$ by $z=4.9$.}. The
$T$--$\Delta$ relation is varied over $\log(T_{0}/\rm K)=3.6$--$5.0$
and $\gamma=0.6$--$1.8$.  The former range is consistent with
estimates of the IGM temperature at mean density at $z \simeq 5$,
while the latter encompasses physically plausible values of $\gamma$
\citep{Becker_2011MNRAS,McQuinnSanderbeck2016}.  We apply flat priors
for all the free parameters except for \taueff, where we instead use a
Gaussian prior with mean \taueff$=1.53$ and a $1\sigma$ uncertainty
corresponding to 4 per cent of the mean, based on the observational
measurement from \citet{Becker_2011MNRAS}.  The range of \taueff
values on our grid of simulations are $0.7-1.3$ times the mean
effective optical depth.  If we use a flat rather than Gaussian prior,
we find the recovery of the thermal parameters is degraded by the
freedom to increase (decrease) the amplitude of the power spectrum on
all scales as \taueff\ is increased (decreased).  In total, we have
$9\times13\times7\times7=5733$ grid points in our model parameter
space.  The mock spectra for each parameter combination on this grid
of models are post-processed by convolving with a Gaussian
instrumental profile of $\rm{FWHM}=7\rm \,km\,s^{-1}$ and rebinning to
$3\rm \,km\,s^{-1}$ per pixel. Gaussian distributed noise is added and
\taueff\ is rescaled iteratively to match the required value.  Once
the model \Lya forest power spectrum parameters are selected,
$\emph{P}_{\rm{F}}^{\rm{model}}$ is obtained by performing a
multi-linear interpolation on the grid of models.

We match the binning of the \Lya forest power spectrum to mock
observations, $\emph{P}_{\rm F}^{\rm data}$, that we extract from one
of our simulations.  These consist of $20$ data points equally spaced
in $\log(k/\rm km^{-1}\,s)$.  We consider two simple data scenarios,
which we describe as ``realistic'' and ``optimistic''.  The former is
comparable to existing \Lya forest data sets at $z\simeq 5$
\citep{Becker_2015_GP}, while the latter may be more appropriate for
observations with high resolution spectrographs on $30$ metre class
telescopes in the forthcoming decade \citep[e.g.][]{Maiolino2013}.  In
the realistic case, we consider a total redshift path length of
$\Delta z=4$, a signal-to-noise ratio $\rm S/N=15$ per pixel and bin
the power spectrum over the range $-2.3<\mathrm{log(k}/$\kmsi)$<-0.7$.
For the optimistic case, we instead adopt a redshift path length five
times larger, $\Delta z=20$, and a higher signal-to-noise per pixel,
$\rm S/N=50$.  The significantly higher signal-to-noise allows the
power spectrum to be measured to smaller scales, up to a maximum
wavenumber of $\mathrm{log(k}/$\kmsi)$=-0.5$.  As demonstrated earlier
in Fig.~\ref{fig:powerspec}, small scale information assists in
breaking the degeneracy between thermal broadening and pressure
smoothing.

We compute the mean and the distribution for each mock data point by
performing $5000$ bootstrap samples with replacement. The covariance
matrix, $\Sigma_{\rm{data}}$, is also determined from these
distributions. As this matrix can be noisy for real data, following
\citet{Lidz_2006ApJ} and \citet{Viel_2013PhRvD} we regularise the
covariance matrix using the correlation coefficients obtained from all
1000 sight-lines drawn from each simulation.  Finally, we increase the
$1\sigma$ bootstrapped uncertainties by $30$ per cent to account for a
possible underestimate in the sample variance
\citep{Rollinde_2013MNRAS} and invert the matrix using singular value
decomposition.  For each mock observation, $\emph{P}_{\rm F}^{\rm
  data}$, we perform $10^{6}$ Markov chain iterations and discard the
first half of the chain as the burn-in.  We verify all chains are
converged by visual inspection.

\subsection{Distinguishing between reionisation models with $P_{\rm{F}}\rm{(k)}$}\label{sec:recovery}

Table~\ref{tab:mcmc} summarises the results of our MCMC analysis for
the realistic and optimistic scenarios for a selection of our models
(for $\log T_{0}$ and $u_{0}$ only), and Fig.~\ref{fig:cpD15} displays
the predicted parameter constraints for the D15 model.

In general we find the model parameters are recovered accurately, with
only a few exceptions that we shall discuss below.  As was
(qualitatively) apparent from Fig.~\ref{fig:powerspec}, we find the
power spectrum is rather insensitive to the slope of the $T$--$\Delta$
relation. The parameter $\gamma$ is recovered within the 68 per cent
credible interval but with fairly broad bounds for most of our models,
even for the optimistic data set. Fig.~\ref{fig:cpD15} indicates it
will be difficult to obtain precise constraints on this parameter from
the \Lya power spectrum alone at $z \simeq 5$, although probing gas at
somewhat higher densities with a joint analysis of the \Lyb forest may
improve this situation
\citep{Dijkstra2004,Furlanetto_2009ApJ,Irsic2014,Boera_2016MNRAS}. On
the other hand, in the absence of significant systematics it should be
possible to jointly constrain $T_{0}$ and \uo using existing \Lya
forest data at $z \simeq 5$ when including the power spectrum on
scales, $\log(k/\rm km^{-1}\,s)>-1$.  Our MCMC analysis indicates that
with current data, the cumulative energy deposited per proton at mean
density may be constrained to a statistical precision of around $\sim
20$ per cent, corresponding to the 68 per cent credible interval.  The
optimistic data scenario instead yields $\sim 8$ per cent, again at
the 68 per cent credible interval.  However, as we discuss in the next
section, systematic uncertainties from observational and numerical
effects will also be important to consider.

\begin{table*}
\caption{Predicted constraints on $\rm{log(}T_0)$ and \uo\ obtained
  from mock observations for the realistic and optimistic data
  scenarios (see text for details). From left to right, the columns
  list the simulation used to construct the mock observation, the
  parameters used in the simulation and the predicted constraints. The
  values correspond to the median of the marginalised posterior
  distribution for each parameter, along with the 68 and 95 per cent
  credible intervals.  The final two rows correspond to the
  constraints from the mock data after an additional $20$ per cent
  systematic uncertainty in the transmitted flux power spectrum at all
  scales is added in quadrature to the bootstrap error bars (see text
  for details).}
\label{tab:mcmc}
\begin{tabular}{cccccccc}
\hline
\hline
&\multicolumn{2}{c}{Model values} & \multicolumn{2}{c}{``Realistic'' scenario} & \multicolumn{2}{c}{``Optimistic'' scenario}\\
\hline  & $\rm{log(T}_0/K)$ & $u_0$\evp& $\rm{log(T}_0/K)$ & $u_0$\evp& $\rm log(T_0/K)$ & $u_0$\evp    
\\Model& & $4.9\leq z\leq 11.5$ & $68\%$ ($95\%$) C.I. & $68\%$ ($95\%$) C.I. &$68\%$ ($95\%$) C.I. &$68\%$ ($95\%$) C.I. &\\\hline        
B15 & $3.98$ & $5.9$ &                                                               
${3.96}_{-0.07}^{+0.08}$ (${}_{-0.14}^{+0.15}$) & ${6.1}_{-1.1}^{+1.0}$ (${}_{-2.2}^{+1.9}$) & 
${3.97}_{-0.03}^{+0.03}$ (${}_{-0.05}^{+0.05}$) & ${6.1}_{-0.5}^{+0.4}$ (${}_{-1.0}^{+0.9}$) &\\                                                        
D15 &  $4.28$ & $11.5$ &                                                              
${4.27}_{-0.06}^{+0.06}$ (${}_{-0.12}^{+0.10}$) & ${12.3}_{-1.7}^{+2.0}$ (${}_{-3.2}^{+4.0}$) &
${4.28}_{-0.02}^{+0.02}$ (${}_{-0.04}^{+0.03}$) & ${11.8}_{-0.8}^{+0.9}$ (${}_{-1.4}^{+1.9}$) &\\
F15 &  $4.47$ & $17.8$ &                                                              
${4.47}_{-0.04}^{+0.04}$ (${}_{-0.08}^{+0.07}$) & ${18.2}_{-2.2}^{+1.7}$ (${}_{-4.0}^{+2.5}$) &
${4.47}_{-0.02}^{+0.02}$ (${}_{-0.04}^{+0.03}$) & ${18.0}_{-1.3}^{+1.3}$ (${}_{-2.4}^{+2.4}$) &\\
Tz12 & $3.93$ & $8.2$ & 
${3.95}_{-0.07}^{+0.06}$ (${}_{-0.13}^{+0.11}$) & ${7.8}_{-0.9}^{+1.0}$ (${}_{-1.7}^{+2.1}$) &
${3.97}_{-0.04}^{+0.03}$ (${}_{-0.08}^{+0.04}$) & ${7.5}_{-0.5}^{+0.6}$ (${}_{-0.9}^{+1.2}$) &\\
Tz9 &$3.92$ & $5.2$ &
${3.90}_{-0.08}^{+0.09}$ (${}_{-0.15}^{+0.16}$) & ${5.6}_{-1.0}^{+0.9}$ (${}_{-2.0}^{+1.7}$) &
${3.90}_{-0.03}^{+0.03}$ (${}_{-0.06}^{+0.05}$) & ${5.6}_{-0.4}^{+0.4}$ (${}_{-0.8}^{+0.8}$) &\\
Tz7 & $3.93$ & $3.7$ & 
${3.89}_{-0.09}^{+0.07}$ (${}_{-0.18}^{+0.12}$) & ${4.2}_{-0.7}^{+0.9}$ (${}_{-1.1}^{+1.8}$) &
${3.96}_{-0.02}^{+0.02}$ (${}_{-0.04}^{+0.04}$) & ${3.5}_{-0.3}^{+0.3}$ (${}_{-0.5}^{+0.6}$) &\\
Tz9HOT & $4.21$ & $11.3$&
${4.15}_{-0.07}^{+0.06}$ (${}_{-0.12}^{+0.11}$) & ${14.2}_{-1.7}^{+2.0}$ (${}_{-3.0}^{+4.0}$) &
${4.15}_{-0.04}^{+0.03}$ (${}_{-0.08}^{+0.06}$) & ${14.1}_{-0.9}^{+1.0}$ (${}_{-1.6}^{+2.1}$) &\\

\hline
D15$+$sys. &  $4.28$ & $11.5$ &                                                              
${4.28}_{-0.09}^{+0.10}$ (${}_{-0.17}^{+0.19}$) & ${12.4}_{-3.1}^{+3.3}$ (${}_{-6.4}^{+6.5}$) &
${4.27}_{-0.05}^{+0.05}$ (${}_{-0.10}^{+0.11}$) & ${12.9}_{-2.5}^{+3.0}$ (${}_{-4.7}^{+6.1}$) &\\

Tz9$+$sys. &$3.92$ & $5.2$ &
${3.90}_{-0.12}^{+0.13}$ (${}_{-0.20}^{+0.22}$) & ${5.8}_{-1.5}^{+1.5}$ (${}_{-2.5}^{+3.1}$) &
${3.92}_{-0.06}^{+0.06}$ (${}_{-0.12}^{+0.13}$) & ${5.5}_{-1.1}^{+1.3}$ (${}_{-2.1}^{+2.7}$) &\\
\hline
\hline

\end{tabular}
\end{table*}

The analysis also demonstrates that such a measurement should already
be able to distinguish between some reionisation scenarios. The one
dimensional posterior distributions for \uo\ obtained from the Tz12,
Tz9 and Tz7 models are displayed in Fig.~\ref{fig:prob_uo}. Recall
that these models have $T$--$\Delta$ relations which are almost
identical at $z=4.9$, but rather different integrated thermal
histories.  We do not consider the Tz15 model -- as already discussed
the power spectrum for this model is very similar to the Tz12
simulation.  On performing the full MCMC analysis, we recover the
cumulative energy input per proton from $4.9 \leq z \leq 11.5$ in the
simulations to within $1\sigma$, and at a precision comparable to the
results in Fig.~\ref{fig:cpD15}. Note again, however, that the
redshift above which the pressure smoothing scale no longer retains a
memory of the thermal history will be model dependent
\citep[cf.][]{Pawlik_2009MNRAS}.  In addition, we find in this case
the peaks of the posterior distributions do not match exactly to the
true value of the parameters in the simulations. This is because only
the A15--G15 models were used to construct the parameter grid in the
MCMC analysis.

As a further demonstration of the model dependent nature of these
predicted constraints, we also construct mock observations from the
Tz9HOT model where the IGM is heated to around $T\simeq 20\,000\rm\,K$
following reionisation. In Fig.~\ref{fig:cpTz9HOT}, it is clear
the recovered $\log T_{0}$ and \uo\ are only consistent within the 95
per cent credible interval for the realistic scenario.  The smaller
statistical error bars obtained in the optimistic case are now
inconsistent with the 95 per cent credible interval for $u_{0}$.
Clearly, an accurate recovery of the thermal history relies on the
grid of models used within the MCMC procedure. This suggests that
developing a set of numerical models which sample the $u_{0}$--$\log
T_{0}$ parameter space as widely and frequently as is practical
will therefore be vital for measuring these parameters using
observational data.

\subsection{Systematic uncertainties}\label{sec:syserror}

Observational and numerical systematics will also impact on the
recovery of \uo\ from the transmitted flux power spectrum.  These have
already been quantified in detail by \citet{Viel_2013PhRvD} (hereafter
V13) in the context of constraining the mass of a putative warm dark
matter particle at $z\simeq 5$.  However, we also briefly outline
these here for completeness and estimate their contribution to the
total uncertainty budget.

There are four main sources of systematic uncertainty to
consider. Following V13, in approximately ascending order of
importance, these are (i) metal line contamination; (ii) the numerical
convergence of the simulations; (iii) spatial fluctuations in the
ionisation state of the IGM and (iv) continuum placement on the
observational data.  Note the impact of galactic outflows on the \Lya
forest is expected to minimal by $z \ga 4$ \citep{Viel2013feedback}.

Narrow metal absorption lines at $z\simeq 5$ arising from \CIV,
\SiIV\ and \MgII\ at lower redshifts have only a minimal effect ($<1$
per cent) on scales $\log(k/\rm km^{-1}\,s)<-1$ (V13).  However, the
contribution of metals to the power spectrum may become more important
toward smaller scales.  We find data at $\log(k/\rm km^{-1}\,s)>-1$ is
important for breaking the degeneracy between thermal broadening and
pressure smoothing, and metals may impact here at the $\sim 5$ per
cent level.  Corrections to the numerical convergence of the
simulations with mass resolution and box size must be applied to the
simulations from the results of convergence tests. V13 estimate an
additional systematic uncertainty of $\sim 5$ per cent in addition to
this known correction.  Spatial fluctuations in the background
ionisation rate, particularly if the mean free path for Lyman
continuum photons is small and/or the ionising sources are rare
\citep{DaviesFurlanetto2015,Chardin2015}, may have a $\sim 10$ per
cent impact on the power spectrum on the scales of interest here. V13
include this as an additional parameter, $f_{\rm UV}$, which is
marginalised over in their MCMC analysis.  Finally, the placement of
the continuum on high resolution quasar spectra is uncertain at around
$10$--$20$ per cent at $z \simeq 5$, which translates to a comparable
uncertainty on the amplitude of the power spectrum.  In practice, this
uncertainty can be forward modelled in the mock spectra \citep[see
  e.g. V13 and][]{Faucher_2008ApJ}.

We estimate the total systematic uncertainty by adding these
contributions in quadrature, yielding $\sim 15$--$25$ per cent for the
\Lya forest power spectrum on the scales of interest.  We estimate the
effect on the precision of the measurements by adding in quadrature an
additional $20$ per cent uncertainty on $P_{F}(k)$ to our bootstrapped
error bars before performing the MCMC analysis.  The resulting
parameter constraints for the D15 and Tz9 models are displayed in the
last two rows of Table~\ref{tab:mcmc}.  This suggests that
measurements of \uo\ with a total uncertainty of $\sim 28$ (22) per
cent are achievable with the the realistic (optimistic) data
scenarios.  Improving the precision of this measurement substantially
will thus require both higher signal-to-noise data as well as careful
forward modelling of the observational and numerical systematics.

\section{Conclusions and discussion} \label{sec:conclude}

\begin{figure}
\includegraphics[width=\columnwidth,trim=0.5cm 1.3cm 2.0cm 2.5cm, clip=true]{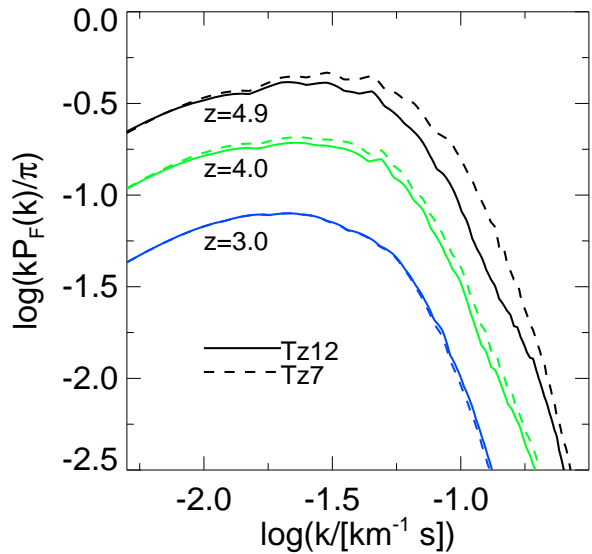}
\vspace{-0.5cm}
\caption{The \Lya forest transmitted flux power spectrum at
    $z=4.9$ (black), $z=4.0$ (green) and $z=3.0$ (blue) for the Tz12
    (solid) and Tz7 (dashed) models. All mock spectra are scaled to
    have $\tau_{\rm eff}=[1.53,0.88,0.39]$ at $z=[4.9,4.0,3.0]$
    \citep{Becker2013_taueff} and have been convolved with a Gaussian
    with FWHM=$7\rm\, {km\,s}^{-1}$.}
\label{fig:powerspecz} 
\end{figure}

In this work we examine the feasibility of constraining the integrated
thermal history at $z>5$ with the \Lya forest using the line of sight
transmitted flux power spectrum. We suggest the cumulative energy
deposited per proton, $u_{0}$, into a gas parcel at mean density at
$5\la z \la 12$ provides a useful parameterisation of the integrated
thermal history in our simulations. We demonstrate this quantity
correlates well with the underlying gas density power spectrum for
$\Delta<10$ over the scales where pressure smoothing acts in the low
density IGM at $z \simeq 5$.  

We also note that $z\simeq5$ observations of the \Lya forest are
well suited for this measurement, despite the fact that most of high
  quality data is available at lower redshifts.  This is demonstrated
  in Fig.~\ref{fig:powerspecz}, which displays the transmitted flux
  power spectrum for the Tz12 and Tz7 models at $z\simeq5,\,4$ and
  $3$.  Recall that both models have very similar instantaneous
  temperatures at mean density, $T_{0}$, at $z<6$ (see
  Fig.~\ref{fig:u_t_sim}).  The differences associated with the
  thermal history at $z>6$ are larger at higher redshift; the models
  are almost indistinguishable by $z=3$ following the response of the
  low density gas to changes in the gas pressure and ongoing Hubble
  expansion.  Furthermore, since \HeII reionisation is expected to
  heat the IGM at $z<5$ \citep[e.g.][]{Becker_2011MNRAS}, higher
  redshift measurements that potentially avoid this additional heating
  are desirable for examining \HI reionisation.

We next perform an MCMC analysis of the transmitted flux power
spectrum using mock observations drawn from a suite of hydrodynamical
simulations. Constraints on the slope of the temperature-density
relation, $\gamma$, are generally weak at $z \simeq 5$.  However, the
degeneracy between thermal broadening and pressure smoothing can be
broken at $z \simeq 5$ using the power spectrum at scales $\log( k/\rm
km^{-1}\,s)>-1$. We estimate \uo\ may be measured with a statistical
uncertainty of $\sim 20$ ($\sim 8$) per cent at $z\simeq 5$ with a
redshift path length of $\Delta z=4$ ($\Delta z=20$) and a typical
signal-to-noise per pixel of $\rm S/N=15$ ($50$) using the power
spectrum to scales $\log( k/\rm km^{-1}\,s)=-0.7$ ($-0.5$). We note,
however, that the constraints are model dependent, and a larger grid
of numerical models which explore the full range of the $u_{0}$--$\log
T_{0}$ parameter space will be required for an in depth analysis of
the observed power spectrum.  Estimates for the expected systematic
uncertainties ($\sim 15$--$25$ per cent) are furthermore comparable to
the statistical precision attainable with current data. Higher
precision measurements are possible only if these systematic
uncertainties are minimised in combination with improved
signal-to-noise and increased path length.

Including systematic uncertainties, we conclude that currently
available data alone should allow for a measurement of \uo\ to within
$\sim 30$ per cent at 68 per cent confidence. This corresponds to
distinguishing between reionisation scenarios with similar
instantaneous temperatures, $T_{0}$, at $z\simeq 5$, but an energy
deposited per proton that varies by $\simeq 2$--$3\, \rm eV$ over the
redshift interval $5\leq z \leq 12$. For an initial $T\sim
10^{4}\rm\,K$ following reionisation, this corresponds to the
difference between early ($z_{\rm re}=12$) and late ($z_{\rm re}=7$)
reionisation in our models.  When compared to predictions of models
for the redshift evolution of the ionising background during
reionisation -- for which $u_{0}$ should be straightforward to compute
-- this will provide an additional and novel constraint on the timing
of the reionisation epoch.

\section*{Acknowledgments}

The hydrodynamical simulations used in this work were performed with
the DiRAC High Performance Computing System (HPCS) and the COSMOS
shared memory service at the University of Cambridge. These are
operated on behalf of the STFC DiRAC HPC facility.  This equipment is
funded by BIS National E-infrastructure capital grant ST/J005673/1 and
STFC grants ST/H008586/1, ST/K00333X/1.  We thank Volker Springel for
making GADGET-3 available. FN is supported by a Vice-Chancellor's
Scholarship for Research Excellence.  JSB acknowledges the support of
a Royal Society University Research Fellowship. We thank the
  anonymous referee for a thoughtful report that helped improve this
  paper.

\bibliographystyle{mnras}
\bibliography{bibliography} 

\label{lastpage}
\end{document}